\documentclass[apj,iop,twocolappendix,numberedappendix]{emulateapj}

\usepackage{natbib}
\usepackage{longtable}
\usepackage{graphicx}
\usepackage[caption=false]{subfig}
\usepackage{float}
\usepackage{epstopdf}
\usepackage{amsmath}
\usepackage{appendix}
\usepackage{hyperref}
\shorttitle{The BH-Bulge Mass Relation with Dwarf Galaxies}
\shortauthors{Schutte et al.}

\begin{document}


\title{The Black Hole - Bulge Mass Relation Including Dwarf Galaxies Hosting Active Galactic Nuclei}


\author{Zachary Schutte}
\affil{eXtreme Gravity Institute, Department of Physics, Montana State University, MT 59715, USA}

\author{Amy E. Reines}
\affil{eXtreme Gravity Institute, Department of Physics, Montana State University, MT 59715, USA}

\author{Jenny E. Greene}
\affil{Department of Astrophysical Sciences, Princeton University, Princeton, NJ 08544, USA}



\begin{abstract}
We present a new relationship between central black hole (BH) mass and host galaxy stellar bulge mass extending to the lowest BH masses known in dwarf galaxies ($M_{\rm BH} \lesssim 10^{5} M_{\odot}$; $M_{\star} \sim 10^{9} M_{\odot}$).  We have obtained visible and near-infrared {\it Hubble Space Telescope (HST)} imaging of seven dwarf galaxies with optically-selected broad-line active galactic nuclei (AGN) and BH mass estimates from single epoch spectroscopy. We perform 2D photometric modeling with GALFIT to decompose the structure of these galaxies and find that the majority have an inner bulge/pseudobulge component with an exponential disk that dominates the total stellar mass. Using the modeling results and color-dependent mass-to-light ratios, we determine the stellar mass of each photometric component in each galaxy. We determine the $M_{\rm BH} - M_{\rm bulge}$ relation using a total of 12 dwarf galaxies hosting broad-line AGNs, along with a comparison sample of 88 galaxies with dynamical BH masses and 37 reverberation-mapped AGNs. We find a strong correlation between BH mass and bulge mass with ${\rm log}(M_{\rm BH}/M_\odot) = (1.24\pm 0.08)~{\rm log}(M_{\rm bulge}/10^{11} M_\odot) + (8.80\pm 0.09)$.  The near-linear slope and normalization are in good agreement with correlations found previously when only considering higher mass systems. This work has quadrupled the number of dwarf galaxies on the BH-bulge mass relation, with implications for BH seeding and predictions for gravitational wave detections of merging BHs at higher redshifts with {\it LISA}.  
\end{abstract}


\keywords{galaxies: active --- galaxies: dwarf --- galaxies: photometry --- galaxies: supermassive black holes}


\section{Introduction}

Scaling relations between supermassive black hole (BH) mass and properties of their host galaxies (e.g., bulge mass, stellar velocity dispersion, infrared luminosity \citep{KH2013} and total stellar mass \citep{Reines2015}) are powerful tools for studying BH and galaxy evolution. While the number of massive galaxies in the local Universe that have been placed on these scaling relations is relatively large, the same cannot be said of dwarf galaxies with stellar masses $M_* \lesssim 10^{9.5} M_{\odot}$. It is particularly difficult to identify and measure the masses of BHs in dwarfs using dynamical methods since the BH sphere of influence can typically only be resolved for very nearby dwarf galaxies, though recent efforts by \citet{nguyen2019DDBH} have used these methods to place improved constraints on BH mass estimates in some nearby low-mass early-type galaxies.  Therefore, searching for active BHs in dwarf galaxies is currently the most productive approach \citep[for a review, see][]{reines_comastri_2016}.

For many years the only dwarf galaxies known to host active galactic nuclei (AGN) were NGC 4395 \citep{filippenko1989,filippenko2003} and POX 52 \citep{Kunth1987,barth2004pox}, both of which were serendipitous discoveries. Subsequently, there were efforts to identify more of these systems using large-scale surveys such as the Sloan Digital Sky Survey (SDSS) \citep{greene2004activeBH,greene2007activeBH,barth2008SDSS}. More recently, \citet{reines2013} used optical spectroscopic signatures from SDSS to identify $>100$ dwarf galaxies with evidence of AGN activity. Though optical spectroscopic diagnostics have been the most productive way to search for these systems, efforts using optical variability \citep{baldassare2018identifying} and radio/X-ray data have also been successful in finding dwarf galaxies that host active BHs \citep[e.g.,][]{2000_DiMatteo_xrayBH,zhang2009census,gallo2010amuse,reines2011actively_Henize,reines2014candidate,ho2016lowxray,pardo2016xray,chen2017hardxray}. Additional candidates have been identified using mid-infrared color diagnostics \citep{satyapal2014discovery,sartori2015search,marleau2017infrared}, however contamination from dwarf starburst galaxies in these samples is significant \citep{2016Hainline_IR}. \par

As the number of dwarf galaxies hosting AGNs continues to increase, it is important to study the host galaxies in detail to investigate which factors may contribute to the presence of an AGN and to place these systems on scaling relations. Determining whether scaling relations hold at the low-mass end has implications for determining the dominant BH formation scenario \citep{volonteri2008evolution,greene2012low,natarajan2014seeds,ricartenatarajan2018}. Additionally, with gravitational waves from massive BH binaries ($10^3 M_{\odot} < M_{binary} < 10^6 M_{\odot}$) being one of the most anticipated targets of \textit{LISA} \citep{amaro2017lisa}, studying dwarf galaxies hosting AGN will help place constraints on the expected detection rates of coalescing massive BH binaries \citep{tamfal2018lisabhb}. \par

With these goals in mind, we present analysis of \textit{Hubble Space Telescope} (\textit{HST}) imaging of seven dwarf galaxies hosting broad-line AGN first identified by \citet{reines2013}.
With our new high resolution \textit{HST} observations, we characterize the structures and morphologies of the dwarf galaxies using photometric modeling techniques. Bulge stellar masses are then estimated using color-dependent mass-to-light ratios. With these results and spectroscopic BH masses in hand \citep{Reines2015}, we place these galaxies on the $M_{\rm BH} - M_{\rm bulge}$ plane with a comparison sample and provide and updated scaling relation.
Throughout this work we assume a standard $\Lambda$CDM cosmology of $H_{0}$ = 70  km  s$^{-1}$ Mpc$^{-1}$ with $\Omega_{\Lambda} = 0.7$ and $\Omega_{M} = 0.3$. We report all magnitudes in the AB system.


\section{{\it HST} Observations of Active Dwarf Galaxies}

We have observed seven dwarf galaxies hosting
broad-line AGN with {\it HST} at optical and near-IR wavelengths (see Table \ref{tab:RGG_PARAM}). These systems were identified in \citet{reines2013} by analyzing the spectra of emission line dwarf galaxies in the NASA-Sloan Atlas (NSA), which is based on the spectroscopic catalog of the SDSS Data Release 8 (DR8, \citet{Aihara_SDSS}). The active dwarf galaxies were identified as broad-line AGNs or Composites using narrow-line diagnostic diagrams (BPT diagram, \cite{1981BPT,Kewley06}) and searching for broad H$\alpha$ emission.  For these broad-line systems, BH masses were estimated using standard virial techniques.  {\it Chandra} observations confirm that our target dwarf galaxies do indeed host massive BHs, as the X-ray luminosities are well above that expected from star-formation-related emission \citep{baldassare2017x}. Throughout this work we refer to these galaxies with the naming scheme set out in \citet{reines2013} in which each galaxy is identified by RGG \# (see Table \ref{tab:RGG_PARAM}).  


\begin{deluxetable*}{lccccccc}
\tablecaption{Dwarf Galaxy Sample}
\tablewidth{7in}
\tablehead{
\colhead{ID} & \colhead{NSA ID} & \colhead{SDSS Name} & \colhead{zdist} & \colhead{Distance (Mpc)} & \colhead{log ($M_{*,\rm total}/M_{\odot})$} & \colhead{log ($M_{\rm BH}/M_{\odot}$)} \\
\colhead{(1)} & \colhead{(2)} & \colhead{(3)} & \colhead{(4)} & \colhead{(5)} & \colhead{(6)} & \colhead{(7)}
}
\startdata
RGG 1 & 62996 & J024656.39$-$003304.8 & 0.0462 & 197.9 & 9.45 & 5.80   \\
RGG 9 & 10779 & J090613.75$+$561015.5 & 0.0469 & 200.9 & 9.30 & 5.44  \\
RGG 11 & 125318 & J095418.15$+$471725.1 & 0.0328 & 140.5 & 9.24 & 5.00   \\
RGG 32 & 15235 &  J144012.70$+$024743.5 & 0.0295 & 126.3 & 9.30 & 5.29 \\
RGG 48 & 47066 &  J085125.81$+$393541.7 & 0.0411 & 176.0 & 9.12 & 5.42 \\
RGG 119 & 79874 &  J152637.36$+$065941.6 & 0.0382 & 163.6 & 9.36 & 5.79 \\
RGG 127 & 99052 & J160531.84$+$174826.1 & 0.0317 & 135.8 & 9.36 & 5.21
\enddata
\tablecomments{Column 1: identification number assigned by \citet{reines2013}, used in this paper. Column 2: NSA identification number. Column 3: SDSS name. Column 4: redshift (zdist) provided in the NSA catalog. Column 5: distance to galaxy in Mpc, determined from redshift given in Column 4 with a Hubble constant of $H_{0}$ = 70  km  s$^{-1}$ Mpc$^{-1}$. Column 6: total stellar mass of the host galaxy based on SDSS magnitudes, computed by \citet{Reines2015}. Column 7: black hole mass as computed by \citet{Reines2015}.}
\label{tab:RGG_PARAM}
\end{deluxetable*}


The {\it HST} images were taken with the Wide Field Camera 3 (WFC3) during February to June 2015 (Proposal 13943, PI: Reines).  One orbit was allocated per galaxy and images were taken with the UVIS F606W filter and the IR F110W filter\footnote{UVIS F275W observations were also taken and are presented in \citet{baldassare2017x}.}. These filters correspond to a wide $V$ and $J$ band respectively. A four point dither pattern was employed for the IR images while a three point pattern was used for the UVIS images. The images were processed using the AstroDrizzle routine in the DrizzlePac software employed by the STScI data reduction pipeline. The native pixel scales (0.04"/pix for the UVIS channel and 0.13"/pix for the IR channel) were preserved in the drizzling process.


\section{Photometric Fitting}

To study the structures and morphologies of our sample galaxies, we fit the 2-D light profiles using GALFIT \citep{peng2002,peng2010}. Of the many analytic functions GALFIT offers we chose to use the general Sersic \citep{Sersic63} profile, which has the functional form 

\begin{equation}
\Sigma (r) = \Sigma_{e} \exp \left[-\kappa \left( \left( \frac{r}{r_e} \right) ^{\frac{1}{n}} - 1 \right) \right],
\end{equation}

\noindent
where $\Sigma_e$ is the pixel surface brightness at radius \textit{r$_e$} and the shape of the profile is determined by the Sersic index \textit{n}. With this versatile function an exponential disk can be modeled with n = 1. Classical bulge components are usually modeled with n \textgreater 2, with the de Vacouleurs profile \citep{Devac1948} the case of n = 4. Other photometric features, such as bars, are typically fit with n $\approx$ 0.5, which is a Gaussian profile. \par

\subsection{Fitting Methodology}

Before fitting the 2-D light profiles of our target galaxies in GALFIT, we created point-spread functions (PSFs) to model the unresolved light coming from the central AGNs.  We used StarFit \citep{Starfit_Hamilton}, which accounts for changes in the PSF due to telescope breathing and the changing position of sources on the detector between different observations. StarFit creates a TinyTim PSF model \citep{krist1995simulation} and matches the focus of the telescope by fitting the model to a source in the observation field (in our case the central bright pixel of each galaxy). This process is performed on each individual frame in the dither pattern and the resulting PSF models are drizzled together with the same parameters as the observations. We found the PSFs generated in this manner worked well for our purposes, matching the increase in central light from the AGN accurately and allowing the other components to be fit freely. We used Starfit to create the PSFs for images taken in the F110W and F606W filters for each galaxy. \par


\begin{deluxetable*}{lcccccccc}

\tablewidth{7in}
\tablecaption{Galfit Model Parameters (F110W)}
\tablecomments{Column 1: identification number used in this paper. Column 2: basic components of the GALFIT model. Column 3: total AB magnitude of each model component reported by GALFIT for the F110W filter. Column 4: Sersic index of each component. Column 5: Half light radius of each component. Column 6: axis ratio (b/a) for each Sersic component. Column 7: position angle of each Sersic component, measured in degrees East of North. Column 8: reduced $\chi^2$ for the best fit model in GALFIT. Column 9: Additional components included in GALFIT model.}
\tablehead{
\colhead{ID} & \colhead{Component} & \colhead{$m_{\rm F110W}$} & \colhead{n}  & \colhead{$R_e$} & \colhead{q} & \colhead{PA} & \colhead{$\chi^2_{\nu}$} & \colhead{Additional} \\
\colhead{} & \colhead{} & \colhead{} & \colhead{} & \colhead{(kpc)} & \colhead{} & \colhead{($^\circ$E of N)} & \colhead{} & \colhead{Components} \\
\colhead{(1)} & \colhead{(2)} & \colhead{(3)} & \colhead{(4)} & \colhead{(5)} & \colhead{(6)} & \colhead{(7)} & \colhead{(8)} & \colhead{(9)}}
\startdata
\\

 & PSF & 21.86$\pm$0.02 & - & - & - & - &  &   \\
RGG 1 & Inner & 19.52$\pm$0.23 & 0.32$\pm$0.09 & 0.71$\pm$0.03 & 0.51 & 25.48 & 1.043 & -  \\
 & Outer & 17.41$\pm$0.06 & 0.83$\pm$0.13 & 1.61$\pm$0.03 & 0.72 & 25.17 &  &   \\ \\
 
 & PSF & 19.80$\pm$0.06 & - & - & - & - &  &   \\
RGG 9 & Inner & 17.01$\pm$0.04 & 2.30$\pm$0.11 & 1.21$\pm$0.26 & 0.86 & 0.26 & 3.283 & -  \\ \\
 
 & PSF & 19.61$\pm$0.03 & - & - & - & - &  &   \\
RGG 11 & Inner & 18.40$\pm$0.23 & 2.40$\pm$0.18 & 0.13$\pm$0.02 & 0.96 & -90.73 & 1.484 & -  \\
 & Outer & 16.22$\pm$0.18 & 1.69$\pm$0.09 & 2.57$\pm$0.30 & 0.78 & -19.39 &  &   \\ \\
 
  & PSF & 18.45$\pm$0.03 & - & - & - & - &  &   \\
RGG 32 & Inner & 17.77$\pm$0.25 & 1.62$\pm$0.20 & 0.29$\pm$0.02 & 0.90 & -14.13 & 1.091 & -  \\
 & Outer & 16.07$\pm$0.10 & 0.74$\pm$0.03 & 2.03$\pm$0.01 & 0.95 & -72.75 &  &   \\ \\
 
  & PSF & 21.55$\pm$0.01 & - & - & - & - &  &   \\
RGG 48 & Inner & 19.75$\pm$0.07 & 0.61$\pm$0.08 & 0.29$\pm$0.06 & 0.42 & -33.85 & 1.363 & -  \\
 & Outer & 16.64$\pm$0.06 & 0.29$\pm$0.01 & 2.12$\pm$0.30 & 0.48 & -30.68 &  &   \\ \\
 
  & PSF & 18.94$\pm$0.07 & - & - & - & - &  &   \\
RGG 119 & Inner & 19.36$\pm$0.21 & 2.55$\pm$0.47 & 0.17$\pm$0.01 & 0.46 & 6.18 & 1.798 & -  \\
 & Outer & 17.23$\pm$0.05 & 0.91$\pm$0.06 & 1.02$\pm$0.01 & 0.78 & -85.42 &  &   \\ \\

  & PSF & 19.94$\pm$0.01 & - & - & - & - &  &   \\
RGG 127 & Inner & 20.40$\pm$0.01 & 0.95$\pm$0.48 & 0.09$\pm$0.02 & 0.53 & -25.43 & 1.737 & Bar (m$_{\rm bar}^{\rm F110W}$ = 17.75) \\
 & Outer & 18.13$\pm$0.05 & 0.70$\pm$0.23 & 1.25$\pm$0.02 & 0.68 & -32.77 &  &   \\

\enddata
\label{tab:GALFIT_PARAM}
\end{deluxetable*}


We approach the fitting process by initially modeling galaxies in the F110W images with a PSF to account for the light from the AGN and a single Sersic component for the galaxy, in addition to a flat background sky. With this initial model we were able to obtain reasonable estimates for the magnitudes of the PSF components and the sizes of the galaxies. During this initial modeling step we also account for light from foreground stars and other objects in the images that are located close to the target galaxy. Dim objects are fit with a single Gaussian component to remove any light contamination to the galaxy of interest. In the case of RGG9, there is a foreground star that is too bright to be robustly accounted for using a Gaussian component. In this case we create a masked region following the procedure presented in the GALFIT manual \citep{peng2010}, which adequately removes the excess light and allows for robust modeling of the galaxy. 

We then applied a PSF, inner Sersic and outer Sersic model to each galaxy drawing on the information from the simpler model. We find that a three component decomposition results in a significantly better fit than a two component decomposition for five of the seven galaxies in our sample (e.g., $\chi^2_{\nu}$ reduced by $\sim 20\%$ or more).  RGG 9 is the first exception with a single Sersic component and a PSF providing an adequate fit ($\chi^2_{\nu} = 3.283 $). The second exception is RGG 127, which requires the inclusion of a bar (modeled with an additional Sersic component with $n\sim 0.5$). Inclusion of a bar allows for much cleaner residuals and an acceptable $\chi^2_{\nu}$ of 1.737. 

In the case of RGG 48, the resolution of the F110W filter is not high enough for GALFIT to consistently settle on a three component model. This is due to the presence of many asymmetric structures (e.g., spiral arms, stellar ring, bar) that are not sufficiently resolved in the F110W image, resulting in GALFIT models which are not robust. We therefore develop our GALFIT model for this system using the F606W image with superior resolution, which allows GALFIT to converge on a stable three component model. The final GALFIT model parameters are shown in Table \ref{tab:GALFIT_PARAM}. 

\begin{figure*}
\centering
\subfloat{\includegraphics[width=\textwidth]{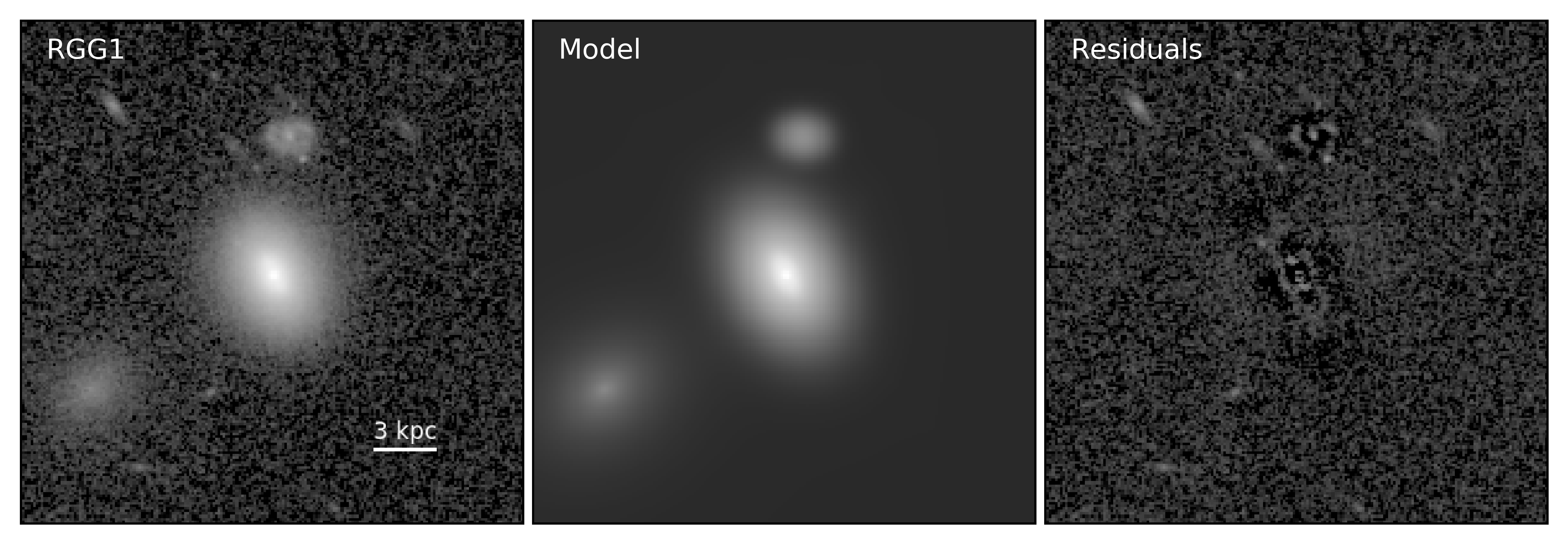}}
\vspace{-0.1cm}
\begin{minipage}{.47\linewidth}
\centering
\subfloat{\includegraphics[scale=0.47]{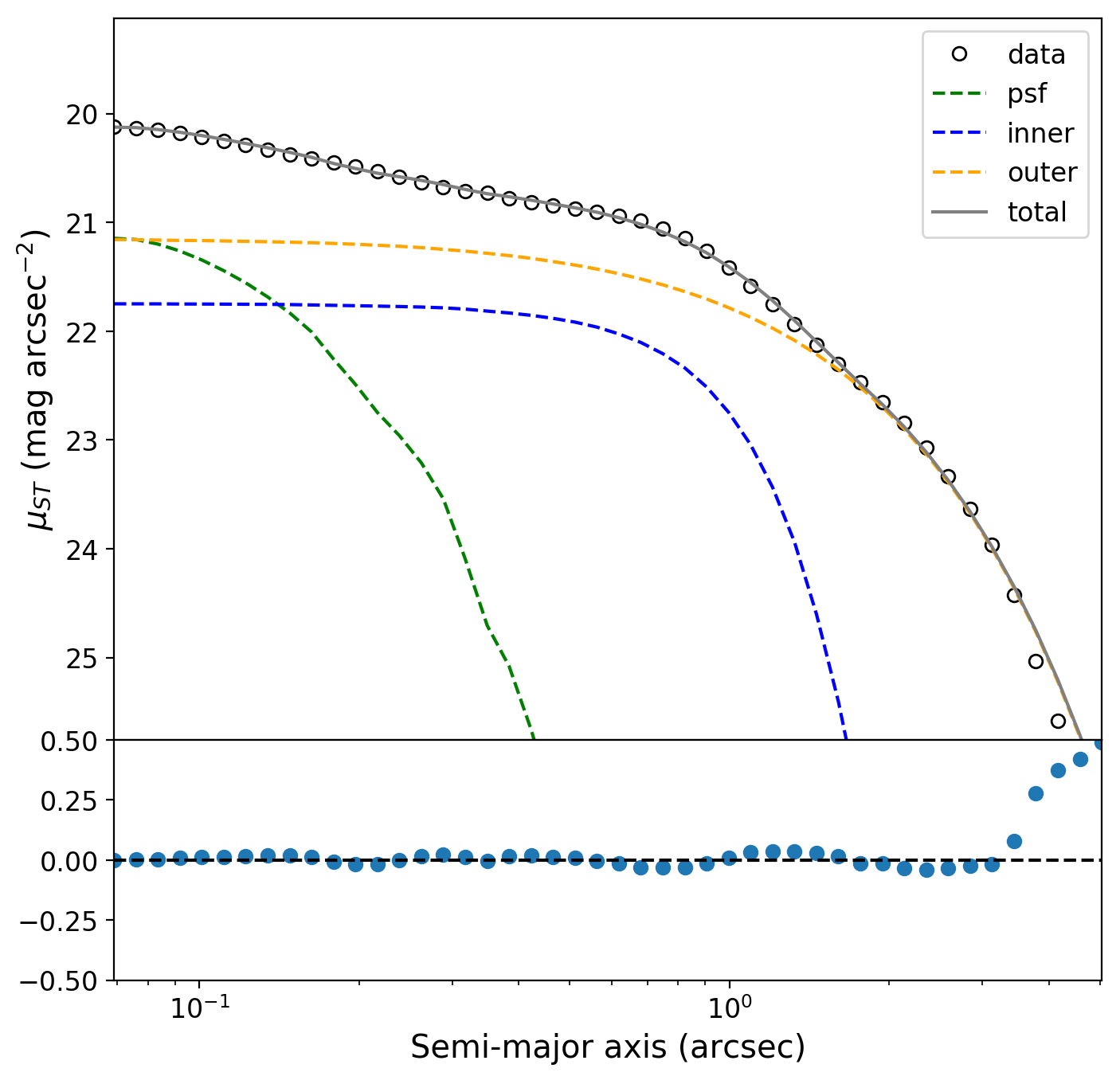}}
\end{minipage}
\hfill
\begin{minipage}{.47\linewidth}
\centering
\subfloat{\includegraphics[scale=0.41]{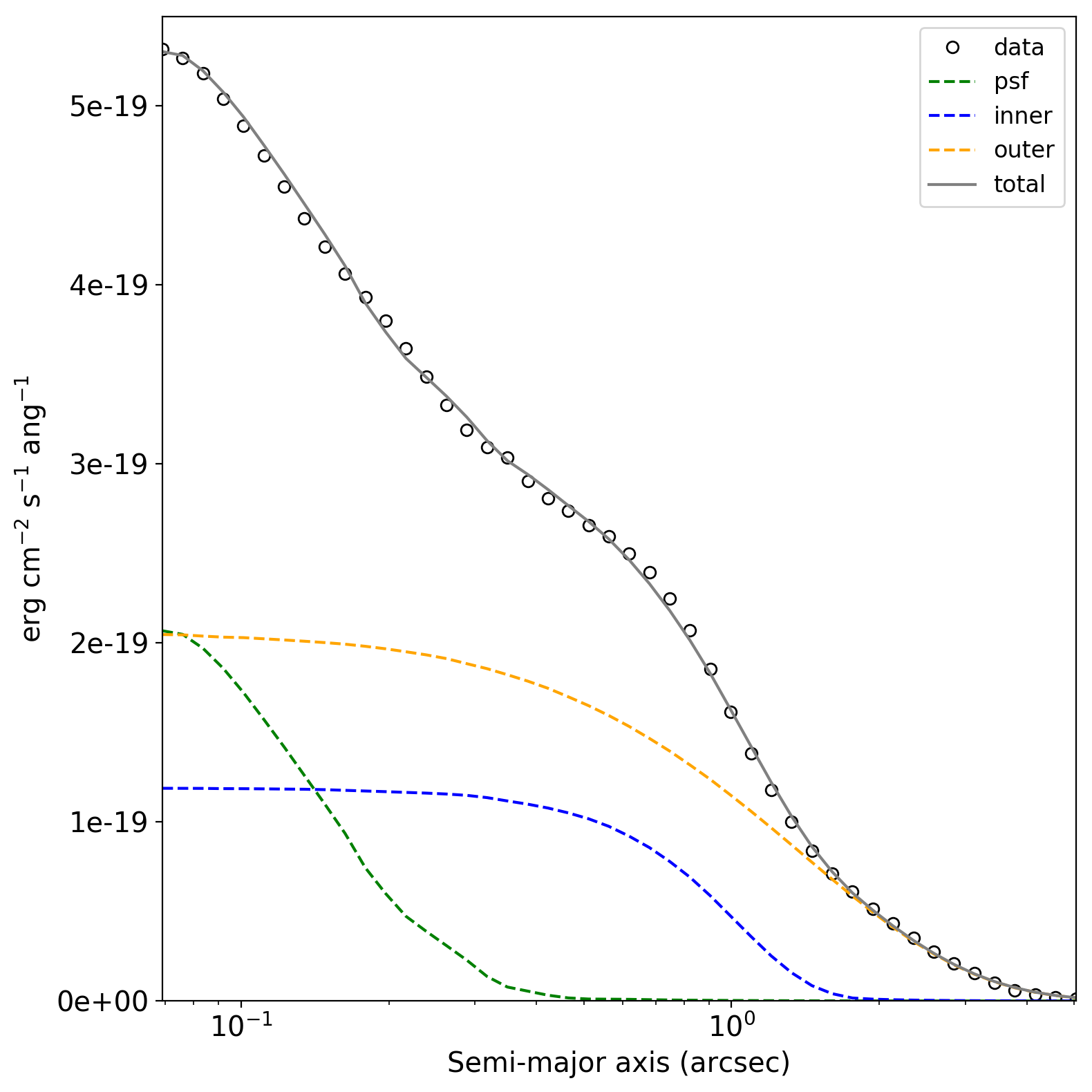}}
\end{minipage}
\caption{Top row: image of RGG 1 in the F110W filter (left); best fit GALFIT model which includes a PSF, inner Sersic component and outer Sersic component (middle); residuals (right). Bottom row: Left panel shows the observed surface brightness profile of RGG 1 with open circles. The best fit model is shown in gray, with the components being shown in green (PSF), blue (inner Sersic) and orange (outer Sersic). The residuals are shown in the lower panel. The right panel shows the average intensity along a given isophote for the data and the intensity as a function of radius. }
\label{fig:RGG1_Model}
\end{figure*}

We also test the necessity of the inclusion of a central point source in our models. To test the robustness of including a PSF, we remove the PSF component from the final model for each galaxy and use these parameters as our initial estimates in GALFIT. This process results in GALFIT being unable to converge to a model in four of the seven systems. For the three systems in which models are achieved the reduced chi squared, $\chi^2_{\nu}$, worsens by an average of 10\%. While this change is somewhat modest, we also observe signatures in the fit residuals (e.g., an overly bright central region surrounded by dark region and bright ring) that indicate the inner Sersic component is converging to a profile which is much too steep near the center. Additionally with the inclusion of a PSF component in our models the fit residuals do not exceed $\sim$ 10$\%$ (residuals can be seen in Figure \ref{fig:RGG1_Model} and the the Appendix), even in the central region of the images where surface brightness rises rapidly and has the largest values. As a final check for PSF necessity we perform a visual inspection of the images for an unresolved nuclear source. In three of the seven galaxies in our sample (specifically RGG 1, RGG 48 and RGG 119) we find that an unresolved nuclear source is discernible by eye in the F110W and F606W images. Finally we find that when the PSF is omitted from our models the inner Sersic index will often diverge or settle to nonphysical values (n \textgreater 10) to account for the rapid increase in surface brightness at the center of the galaxy. The combination of these factors indicates that a PSF component is justified in our models. 

Once our models are developed for each galaxy in our sample, we apply these results to the images taken in the F606W filter (F110W for RGG 48). To accomplish this we fix the structural parameters reported by GALFIT in our original model (e.g., radii, Sersic index, axis ratio), taking into account the differing pixel scales of the two filters.  While the structural parameters are fixed in this step, we allow the magnitudes of each component to vary freely.  This enables us to obtain magnitude measurements for each photometric component in both filters. In general the models derived in the F110W images (F606W for RGG 48) agree well with the results from fitting the F606W images (F110W for RGG 48) and require little modification. Further discussion concerning the morphology of individual galaxies is provided in the Appendix.


\subsection{Uncertainty in GALFIT Parameters}

To estimate the uncertainty of the magnitudes reported by GALFIT we begin by fixing the background sky value in our model. The background sky value is determined by iteratively $\sigma$-clipping (with 10 iterations and $\sigma$=3) the image used in our GALIT models to mask bright features and taking the median value of the resulting masked image. We then fit the image with the fixed background and take the uncertainty in the magnitude of each component to be the difference between the magnitude from the best fit model and the magnitude from the fixed background model. To estimate the error in the Sersic index and effective radii we replace the PSF in our best fit model with an isolated bright star taken from the image used in the modeling. The error is then taken to be the difference between the resulting fit parameters and those from our best fit model.


\begin{deluxetable*}{lccccccccc}
\tablewidth{6.5in}
\tablecaption{Additional Apparent Magnitudes}
\tablecomments{All magnitudes are reported in the AB system. Column 1: identification number. Column 2: total magnitude of PSF component reported by GALFIT in F606W filter. Column 3: total magnitude of inner Sersic component reported by GALFIT in F606W filter. Column 4: total magnitude of outer Sersic component reported by GALFIT in F606W filter. Column 5: total magnitude of inner Sersic component in SDSS r filter. Column 6: total magnitude of outer Sersic component in SDSS r filter. Column 7: total magnitude of inner Sersic component in SDSS z filter. Column 8: total magnitude of outer Sersic component in SDSS z filter. Column 9: total magnitude of inner Sersic component in 2MASS J filter. Column 10: total magnitude of outer Sersic component in 2MASS J filter. Magnitudes reported for SDSS r, SDSS z and 2MASS J filters are not reported by GALFIT but are determined using the power law fitting procedure described in \S\ref{sec:PhotConversion}.}
\label{tab:Photometry}
\tablehead{
\colhead{ID} & \colhead{$m_{\rm psf}^{\rm F606W}$} & \colhead{$m_{\rm inner}^{\rm F606W}$} & \colhead{$m_{\rm outer}^{\rm F606W}$} & \colhead{$m_{\rm inner}^r$} & \colhead{$m_{\rm outer}^r$}
& \colhead{$m_{\rm inner}^z$} & \colhead{$m_{\rm outer}^z$} & \colhead{$m_{\rm inner}^J$} & \colhead{$m_{\rm outer}^J$} \\
\colhead{(1)} & \colhead{(2)} & \colhead{(3)} & \colhead{(4)} & \colhead{(5)} & \colhead{(6)} &
\colhead{(7)} & \colhead{(8)} & \colhead{(9)} & \colhead{(10)} }
\startdata
RGG 1 & 23.11 & 20.66 & 18.39 & 20.58 & 18.32 & 19.95 & 17.78 & 19.40 & 17.30\\
RGG 9 & 20.49 & 17.80 & - & 17.75 & - & 17.31 & - & 16.92 & -\\
RGG 11 & 19.93 & 19.01 & 17.07 & 18.96 & 17.01 & 18.63 & 16.54 & 1833 & 16.13\\
RGG 32 & 18.86 & 18.44 & 16.93 & 18.39 & 16.87 & 18.03 & 16.40 & 17.70 & 15.98\\
RGG 48 & 21.78 & 20.70 & 17.13 & 20.63 & 17.097 & 20.11 & 16.82 & 19.65 & 16.58\\
RGG 119 & 19.59 & 19.75 & 18.18 & 19.72 & 18.11 & 19.51 & 17.59 & 19.31 & 17.13\\
RGG 127 & 21.17 & 21.82 & 18.88 & 21.72 & 18.83 & 20.94 & 18.42 & 20.25 & 18.04
\enddata
\end{deluxetable*}


\subsection{Fitting Results}

\begin{figure*}
\begin{minipage}{.47\linewidth}
\centering
\subfloat{\includegraphics[scale=0.65]{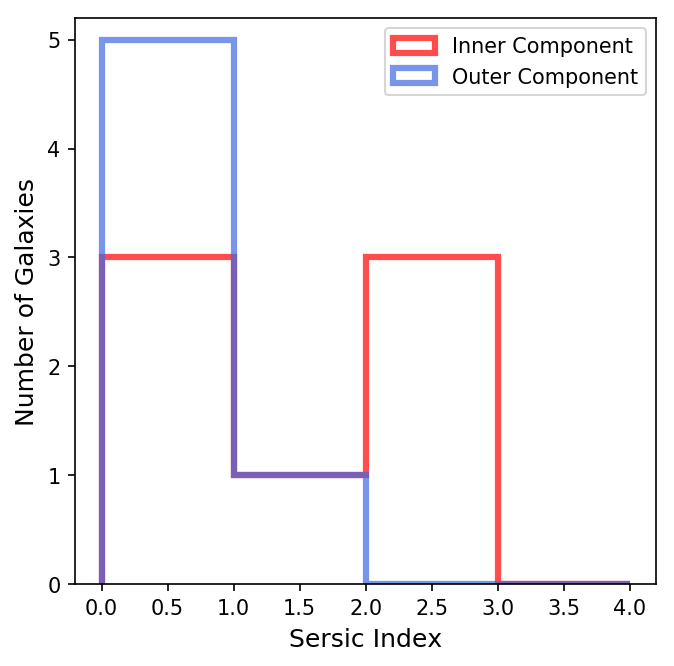}}
\end{minipage}
\hfill
\begin{minipage}{.47\linewidth}
\centering
\subfloat{\includegraphics[scale=0.65]{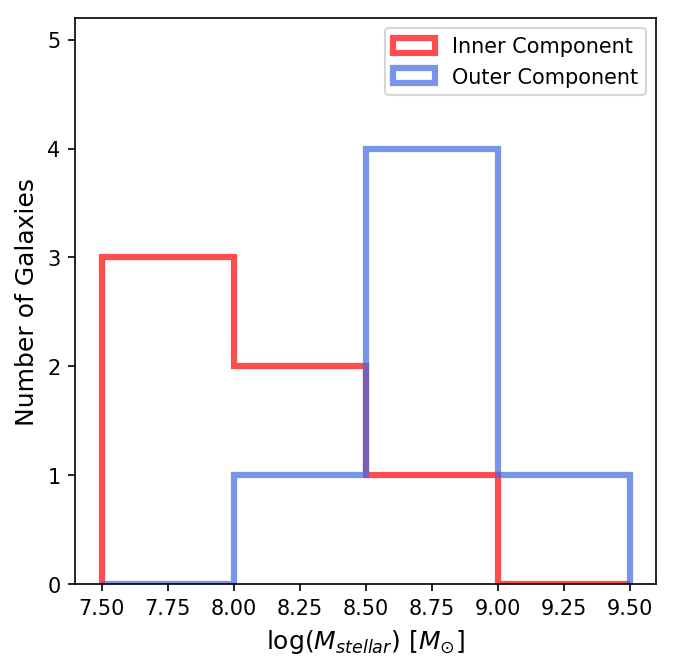}}
\end{minipage}
\caption{Left panel: Distribution of the Sersic index for the inner component and outer component of the best fit models derived with GALFIT \citep{peng2010} using the F110W filter images. Right panel: Distribution of stellar masses for the inner component and outer component. Stellar masses were estimated using the procedure describes in \S\ref{sec:stellarmasses}. }
\label{fig:Param_hist}
\end{figure*}

Overall we find a median inner Sersic index for our sample of 1.6 and a median outer Sersic index of 0.79. The inner Sersic indices span a large range of values (seen in the left panel of Figure \ref{fig:Param_hist}). This indicates that the inner components range from a pseudobulge to a classical bulge morphology (we again refer the reader to \S\ref{sec:gal_decomp} for more detail on the morphological classifications for individual galaxies). The outer Sersic indices show much less variation (left panel of Figure \ref{fig:Param_hist}), and are generally consistent with a Gaussian or exponential disk.  For the entire sample we find that six of the seven systems require two or more Sersic components to produce an acceptable fit. When considering the systems with a detected outer Sersic component we find the ratio between the inner (bulge) component and total light (with AGN contribution excluded) to have a median value of 0.12 and a range from 0.05 to 0.17. These findings are in good agreement with work done by \citet{jiang2011host} who find the median bulge-to-total light ratio to be 0.16 when considering galaxies with a detected disk. \par

To supplement our 2-D GALFIT models we perform elliptical isophotal fitting using the photutils package from the Astropy library \citep{Astropy2013,Astropy2018}. With this we derive 1D radial surface brightness and intensity profiles for each of our galaxies and GALFIT models. An example 
can be seen in Figure \ref{fig:RGG1_Model} for the galaxy RGG 1. The models for the rest of our sample can be found in the Appendix. \par

It is of interest to briefly consider the structural parameters from GALFIT modeling in the context of other non-active dwarf galaxy samples. \citet{amorin2009BCDG} characterize the stellar host structure of 20 blue compact galaxies, using similar 2D Sersic models from GALFIT to model surface brightness profiles. They find that all but one galaxy have low Sersic indexes (0.5 $\lesssim$ n $\lesssim$ 2) and the sample has a mean effective radius of $\sim$ 1.1 kpc. \citet{janz2014VIRGOdwarf} use GALFIT to study the stellar structure of 121 Virgo early type dwarf galaxies using near IR imaging. They perform single and multiple Sersic component fits to find that surface brightness profiles tend to follow an overall exponential shape. This result holds for multiple Sersic fits as well, with the inner component typically having an exponential profile as well (n $<$ 1.2). \citet{lian2015BCDGGOODS} study the surface brightness profiles of 34 blue compact dwarf galaxies found in the Great observatories origins Deep Survey (GOODS) North and South fields from the Cosmic Assembly Near-IR Deep Extragalactic Legacy Survey (CANDELS, \citet{grogin2011candels,koekemoer2011candels}). They perform one and two component Sersic fits to find that approximately half of the galaxies in their sample are better fit with two components. Across the entire sample they find that the effective radius to be less than $\sim$ 4 kpc and that the inner and outer components are well fit by low Sersic indices (n $\lesssim$ 1.5). We find similar structural parameters for our sample of active dwarf galaxies. First, a large fraction of the samples require two or more components to provide adequate fits to the surface brightness profile and the inner component of the multi-component fits often have a low Sersic index (n $\sim$ 1). Additionally, the size of the galaxies across all the samples is relatively consistent, with the effective radius of the outer Sersic component being approximately 1 $\lesssim r_e \lesssim$ 4 kpc. 


\section{Stellar and Black Hole Masses} \label{sec:Masses}

As our GALFIT models provide magnitudes for each photometric component (e.g., bulge/pseudobulge, disk, bar) in two filters for every galaxy, we are able to use these results to estimate stellar masses of each component. This is done by employing color dependent mass-to-light ratios derived by \cite{zibetti2009}. Here we present the process and results found using these relationships. In addition, we also address the single-epoch spectroscopic techniques used to estimate the BH mass for each system in the sample.


\subsection{Photometric Conversions} \label{sec:PhotConversion}

The {\it HST} magnitudes derived using the GALFIT models are comparable to a broad $V$ filter (F606W) and a broad $J$ filter (F110W), but are not equivalent to true Sloan or Johnson filters. The color dependent mass-to-light ratio we employ from \cite{zibetti2009} requires Sloan $r-z$ colors and 2MASS $J$ luminosities (see Equation \ref{eqn:ml}). To estimate the magnitudes in the 2MASS $J$, SDSS $r$ and SDSS $z$ filters, we use the flux density measurements reported by GALFIT in the {\it HST} filters and fit a power law in log(f$_{\lambda}$) versus log($\lambda$) space. We then obtain estimates of flux densities in the $J,r,z$ filters by evaluating the power law fit at the appropriate pivot wavelengths \citep{gunn1998sloan,Cohen2MASS2003}. The flux densities are then converted to AB magnitudes to use in the \citet{zibetti2009} relation. This approach is motivated by stellar population models \citep[e.g.,][]{starburst99} that demonstrate that a power law is a good description of a stellar population spectrum redwards of 4000\AA, which is our region of interest. 

As a consistency check, we also investigate the method described in \citet{lasker2016black} where simple stellar population (SSP) models from the PARSEC code \citep{marigo2017new} are used to generate magnitudes for a SSP in several different filter systems. The models results are then used to derive a relationship between the known magnitudes (F110W and F606W) and the desired magnitudes ($r,z$ and $J$). When comparing the two methods we find no significant change in the magnitudes or the stellar masses produced, and adopt the power law fitting method in this work. When converting the derived \textit{r},\textit{z} and \textit{J} apparent magnitudes to luminosities, the distance to each galaxy was estimated using the zdist parameter reported in the NSA that includes the peculiar velocity model of \citet{willick1997homogeneousvel} (see Table \ref{tab:RGG_PARAM}).


\subsection{Stellar Masses} \label{sec:stellarmasses}

With magnitudes determined with our GALFIT models and converted into the correct filters, we use color dependent mass-to-light ratios developed by \citet{zibetti2009} found in Table B1 of their work. Specifically we use $J$-band luminosity as a function $r-z$ color to determine mass-to-light ratios and compute stellar masses of each structural component via the relation

\begin{equation}
{\rm log}(M/L_J) = 1.398(r-z) - 1.271,
\label{eqn:ml}
\end{equation}

\noindent
adopting a solar absolute magnitude of $M_J = 4.54$ in the AB system \citep{cohen2003}. The resulting stellar masses can be found in Table \ref{tab:masslum}. Errors in these stellar mass estimates are expected to be $\sim$ 0.3 dex and will be dominated by uncertainties in stellar evolution \citep{ConroyStellarMass2009}. \par

We choose to use the color dependent mass-to-light ratios from \citet{zibetti2009} as they derive their relations using stellar population synthesis models with revised prescriptions for the TP-AGB evolutionary phase from \citet{marigo2007evolution} and \citet{marigo2008evolution}. Additionally, \citet{zibetti2009} develop their relations taking into account the effect of dust and, more importantly, allow for young stellar populations and include star formation bursts. They argue that on local scales the effects of dust and variation in star formation history cannot be ignored, whereas in work such as \citet{bell2003} who consider global properties, a smooth star formation for the entire galaxy can be employed. As we are investigating the structure and mass distribution in this study, the relations from \citet{zibetti2009} are a natural choice for our purposes. Nevertheless, we also estimate the stellar masses using the relations from \citet{bell2003} when placing them on the $M_{\rm BH} - M_{\rm bulge}$ relation (see \S\ref{sec:MBHScaling}) to obtain a secondary estimate of the best fit relation and to exemplify the effect of the stellar mass estimation method on BH scaling relations. 

When estimating stellar masses in this way, attention must be given to the different initial mass functions (IMFs) used to derive the mass-to-light ratios. \citet{bell2003} use a `diet' Salpeter \citep{SalpeterIMF1955} IMF while the relations in \citet{zibetti2009} are derived using a Chabrier IMF \citep{ChabrierIMF2003}. For our sample of dwarf galaxies, the \cite{bell2003} relations predict stellar masses which have a median increase of 0.74 dex when compared to stellar masses found using the \citet{zibetti2009} relation with the same colors. When the systems studied in \cite{KH2013} (see \S \ref{sec:additional_sys}) are considered in addition to our sample of dwarf galaxies the median increase falls to 0.24 dex. This difference is consistent with findings by \citet{Reines2015}, where masses calculated with $g-i$ colors using the \citet{bell2003} relation are compared to the \citet{zibetti2009} relation, as well as masses taken from the NSA that are computed using the kcorrect code \citep{BRkcorrect}. We reiterate the importance of estimating stellar masses consistently across samples, as the choice of mass-to-light ratios can have a significant impact on stellar mass estimates.   

In this work, we report stellar mass estimates for the inner and outer structural components of our target galaxies using the \cite{zibetti2009} relations (see Table \ref{tab:masslum}). The median stellar mass for the inner Sersic components is $10^{7.97}M_{\odot}$ with a standard deviation of $\sigma_{M_*,inner} = 0.43$ dex (see the right panel of Figure \ref{fig:Param_hist}). Note that RGG 9 which has an inner component mass of $10^{9.0}M_{\odot}$ is modeled with a PSF and single Sersic component, so the inner component mass will be equivalent to the total stellar mass. The outer Sersic components have a median stellar mass of $10^{8.86}M_{\odot}$ and a standard deviation of $\sigma_{M_*,inner} = 0.31$ dex with the distribution of these masses seen in the right panel of Figure \ref{fig:Param_hist}. With stellar masses for each Sersic component determined, we are able to estimate the bulge-to-total stellar mass ratio for the galaxies in our sample. For the systems which require two Sersic components we find the median bulge-to-total stellar mass ratio to be 0.11 with a range from 0.05 to 0.16, indicating the stellar mass in these systems is dominated by the outer component.

When we compare the total mass of these systems to the estimates calculated by \cite{Reines2015} (given in Column 5 of Table \ref{tab:RGG_PARAM}) we find our estimates are smaller by a median value of $\sim$0.3 dex. This arises as the magnitudes from modeling with GALFIT are slightly dimmer than those in the NSA, which are used by \citet{Reines2015} to estimate the total stellar masses for our sample. Given that the uncertainty in the method used to estimate both our stellar masses and the masses reported by \cite{Reines2015} is $\sim$0.3 dex and that the magnitudes we estimate for our sample are slightly dimmer than those used by \citet{Reines2015}, our results are consistent with their findings. 


\begin{deluxetable*}{lcccccc}
\tablewidth{7in}
\tablecaption{Masses and Luminosities}
\tablecomments{Column 1: identification number. Column 2: stellar mass of the inner Sersic (bulge) component. Column 3: stellar mass of the outer Sersic component. Column 4: 2MASS $J$ band luminosity of the inner Sersic component. Column 5: 2MASS $J$ band luminosity of the outer Sersic component. Column 6: ratio of inner Sersic component stellar mass to total stellar mass. Column 7: ratio of inner Sersic component 2MASS $J$ luminosity to total 2MASS $J$ luminosity (not including light from AGN).}
\tablehead{
\colhead{ID} & \colhead{Log$(M_{\rm inner}/M_{\odot})$} & \colhead{Log$(M_{\rm outer}/M_{\odot})$} & \colhead{Log$(L_{\rm inner}/L_{\odot})$} & \colhead{Log$(L_{\rm outer}/L_{\odot})$} & \colhead{$M_{\rm inner}/M_{\rm total}$} & \colhead{$L_{\rm inner}/L_{\rm total}$} \\
\colhead{(1)} & \colhead{(2)} & \colhead{(3)} & \colhead{(4)} & \colhead{(5)} & \colhead{(6)} & \colhead{(7)}
}
\startdata
RGG 1 & 8.21 & 8.93 & 8.66 & 9.50 & 0.16 & 0.13 \\
RGG 9 & 8.96 & - & 9.67 & - & 1 & 1  \\
RGG 11 & 7.97 & 9.03 & 8.78 & 9.66 & 0.08 & 0.12  \\
RGG 32 & 8.14 & 8.98 & 8.95 & 9.64 & 0.13 & 0.17   \\
RGG 48 & 7.87 & 8.73 & 8.45 & 9.67 & 0.11 & 0.06    \\
RGG 119 & 7.49 & 8.80 & 8.53 & 9.40 & 0.05 & 0.12  \\
RGG 127 & 7.75 & 8.11 & 8.00 & 8.88 & 0.09 & 0.05
\enddata
\label{tab:masslum}
\end{deluxetable*}


\subsection{Black Hole Masses} \label{sec:BHmasses}

We obtain virial BH mass estimates for our sample (see Table \ref{tab:RGG_PARAM}) from \citet{Reines2015} (also see \cite{reines2019erratum}) who derive BH masses from broad H$\alpha$ emission detected in SDSS spectroscopy. Estimating BH masses from single-epoch spectroscopy is a commonly used method relying on the assumption that the broad-line region (BLR) kinematics are dominated by the gravity of the central BH. Under this assumption, the BH mass can be estimated by $M_{BH} \propto R\Delta V^2/G$. The average gas velocity is estimated from the width of a broad emission line and the radius of the BLR is taken from the radius-luminosity relation based on reverberation-mapped AGN (e.g.,\cite{vestergaard2006bhmass},\cite{bentz2013low}). The constant of proportionality in estimating BH mass is dependent on the geometry and orientation of the BLR, which are in general unknown. While this parameter is known to vary from system to system \citep{2009bentz_reverberation,2011barth_reverberation} a single proportionality constant is adopted from calibrating reverberation-mapped BH masses to the $M_{BH}-\sigma_*$ relation \citep{2000gebhardtreverberation,greene2005BHmass,park2012lick,ho2014black}. \par

The BH masses estimated by \cite{Reines2015} were found using equation 1 in their work:

\begin{multline}
 \text{log}\left(\frac{M_{\rm BH}}{M_{\odot}}\right) = \text{log} \epsilon + 6.57 + 0.47 \text{log}\left(\frac{L_{\rm H{\alpha}}}{10^{42}~\text{erg s}^{-1}}\right) \\
 + 2.06\text{log}\left(\frac{\rm FWHM_{H\alpha}}{10^3~\text{km s}^{-1}}\right),
\end{multline}

\noindent
which was derived with the methods of \citet{greene2005BHmass} and incorporates an updated radius-luminosity relationship from \citet{bentz2013low}. \citet{Reines2015} adopted $\epsilon = 1.075$, which corresponds to a mean virial factor $\left<f\right> = 4.3$ from \cite{grier2013stellar}\footnote{This is only slightly different from \citet{reines2013}, who adopt $\epsilon = 1$.}. Estimates of BH masses from single-epoch virial methods are very indirect and have uncertainties of $\sim$0.5 dex \citep{2013shen_BHmasserr}.


\section{Additional Systems} \label{sec:additional_sys}

Here we briefly discuss additional systems included in our investigation of the $M_{\rm BH}-M_{\rm Bulge}$ scaling relation. The constants used in all color dependent mass-to-light ratio mass estimations are from \citet{zibetti2009} and are found in Table B1 of their work.


\subsection{Other Dwarf Galaxies with Broad-line AGNs} \label{sec:dwarf_add}

First we consider five additional dwarf galaxies hosting broad-line AGNs.  The dwarf galaxy NGC 4395 \citep{filippenko1989,filippenko2003} hosts a Seyfert 1 nucleus, with its morphology well described by a disk and nuclear star cluster. As this galaxy does not have a well-defined bulge/pseudobulge component we do not include it when fitting a linear regression to the data (see \S\ref{sec:MBHScaling}). We do, however, place this system on the plot of BH mass vs.\ bulge mass using the total stellar mass from \citet{Reines2015} as an upper limit, and the reverberation-mapped BH mass estimate ($M_\star^{\rm NGC 4395} = 10^{8.90} M_{\odot}$ and $M_{\rm BH}^{\rm NGC 4395} = 10^{5.45} M_{\odot}$). Though \cite{woo2019NGC4395} recently performed an updated reverberation mapping study of NGC 4395 to find a revised BH mass estimate of $M_{\rm BH}^{\rm NGC 4395} = 10^{3.96} M_{\odot}$, we choose to use the larger reverberation mapping mass as it agrees with previous kinematic BH mass estimates. Additionally we include the dwarf Seyfert 1 galaxy POX 52 \citep{barth2004pox,thornton2008} which has no detected disk component and a Sersic index of $n=4.0$. For POX 52 we estimate the stellar mass using the photometry provided by \citet{barth2004pox}, specifically using the relation

\begin{equation}
\text{log} \left(M/L_I\right) = 1.475(B-V) - 1.003
\end{equation}

\noindent
with a solar absolute I band magnitude of 4.10 \citep{mann2015revised}. We find a stellar mass of $M_\star^{\rm POX 52} = 10^{9.26} M_{\odot}$, which is in good agreement with the mass found by \citet{thornton2008} of $M_\star^{\rm POX 52} = 10^{9.08} M_{\odot}$. We adopt the BH mass reported in \citet{thornton2008} for POX 52 of $M_{\rm BH}^{\rm POX 52} = 10^{5.48} M_{\odot}$. \par

We also add the two remaining dwarf galaxies hosting broad-line AGN from \cite{reines2013}, RGG 20 and RGG 123 (SDSS IDs: J122342.81+581446.1 and J153425.59+040806.7 respectively). These systems were previously identified by \citet{greene2007activeBH} and the host galaxies were studied in detail by \cite{jiang2011host}. They perform morphological decompositions using GALFIT on {\it HST}/WFPC2 images taken in the F814W ($\sim I$-band) filter. To obtain bulge stellar mass estimates for these systems, we first take the bulge-to-total luminosity ratio calculated by \cite{jiang2011host} (0.93 for RGG 20 and 0.12 for RGG 123) and assume this to be constant across the SDSS $r,i$ and $z$ filters.  
This allows us to take the Petrosian magnitudes from SDSS (the recommended magnitudes for estimating the total magnitude from an extended source such as a galaxy) and multiply the corresponding flux densities by the bulge-to-total ratio to obtain an estimate for the bulge component in these SDSS filters. With the scaled SDSS magnitudes we calculate bulge stellar masses using the relation,

\begin{equation}
\text{log} \left(M/L_i\right) = 1.797(r-z) - 1.238
\end{equation}

\noindent
with a solar absolute $i$-band magnitude of 4.53 mag \citep{gunn1998sloan}. Using this relation we find the stellar bulge masses of RGG 20 and RGG 123 to be $10^{8.97}M_{\odot}$ and $10^{8.11}M_{\odot}$ respectively. We emphasize that these bulge mass estimates are not as robust as the other RGG dwarf galaxies with new {\it HST} observations, though their inclusion in the BH-bulge mass relation does not effect the outcome when calculating a best fit linear regression (see \S\ref{sec:MBHScaling}). To differentiate these systems from the other dwarf galaxies when placing them on the scaling relation they are plotted with triangles as opposed to stars (see Figure \ref{fig:scaling}). \par

The final dwarf galaxy we include is RGG 118 \citep{reines2013,baldassare2015,baldassare2017}, a spiral system hosting one of the smallest nuclear BHs yet found with a mass of $\sim 50,000~M_{\odot}$. For RGG 118, we use the F160W luminosity and the F475W$-$F775W color (which are equivalent to 2MASS $H$-band luminosity and SDSS $g-i$ color respectively) provided in \citet{baldassare2017} to estimate the stellar mass with the following relation:

\begin{equation}
\text{log} \left(M/L_H\right) = 0.780(g-i) - 1.222
\end{equation}

\noindent
with a solar absolute F160W magnitude of 4.60 mag \citep{cohen2003}. This gives a stellar mass for the bulge component of RGG 118 of $\sim 10^{8.25} {M}_{\odot}$. Our estimate is lower than the bulge stellar mass reported by \cite{baldassare2017} of $\sim 10^{8.59} {M}_{\odot}$ but in good agreement as they use the color dependent mass-to-light ratios from \citet{bell2003} which are expected to predict a larger mass.


\subsection{Galaxies with Dynamical BH Masses} \label{sec:DDBHS}

\citet{KH2013} compile a sample of BH mass measurements which come from a variety of dynamical methods (stellar dynamics, CO molecular gas disk dynamics, ionized gas dynamics and maser disk dynamics). We use 79 of these galaxies in our comparison sample for investigating the $M_{\rm BH}-M_{\rm Bulge}$ relation which enables us to span the entire known mass range of nuclear BHs ($M_{\rm BH} \sim 10^5-10^{10} M_\odot$). 

To estimate the stellar mass of the bulge (or pseudobulge) component for galaxies with dynamical BH masses, we use the absolute $K$-band magnitudes and $B-V$ colors provided by \citet{KH2013}. We include all objects found in \cite{KH2013} except those with BH mass upper limits (2 ellipticals and 2 spiral galaxies with pseudobulges) and galaxies without provided $B-V$ colors which are required for estimating stellar masses (2 additional ellipticals an 3 spiral galaxies with pseudobulges). We use the following relation \citet{zibetti2009}

\begin{equation}
\text{log}\left(M/L_K\right) = 1.176(B-V) - 1.390
\end{equation}

\noindent
to estimate stellar masses for this sample, assuming a solar absolute $K$-band magnitude of 3.32 mag \citep{bell2003}. As \cite{KH2013} report only the integrated $B-V$ colors for the entire galaxy it should be noted that there will be some bias toward lower mass estimates for disk galaxies in their sample arising from bluer colors of disks, which results in a smaller mass-to-light ratio. \par

As discussed in \cite{Reines2015}, \cite{KH2013} estimate $M/L_K$ as a function of $B - V$ color differently. The relation they derive is based on the mass-to-light ratio calibrations of \cite{into2013MLratio}. Using the method of \cite{KH2013} results in stellar masses that are systematically larger than masses estimated using the \cite{zibetti2009} relation by 0.33 dex \citep{Reines2015}. We adopt stellar masses using the \cite{zibetti2009} relations to maintain consistency between all samples used in this work. \par

In addition to the sample of dynamical BH mass measurements from \citet{KH2013}, we also include the galaxies studied by \citet{lasker2016black}. \citet{lasker2016black} perform detailed photometric structure decompositions using \textit{HST} imaging for nine megamaser disk galaxies, yielding luminosities and colors for the identified bulge components. We use these nine late-type galaxies in our comparison sample, estimating the stellar mass using the following relation from \cite{zibetti2009}

\begin{equation}
\text{log}\left(M/L_H\right) = 0.780(g-i) - 1.222
\end{equation}

\noindent
using the $g-i$ colors and $H$-band luminosities provided in Tables 5 and 7 respectively by \cite{lasker2016black}.\par

It should be noted that since the work of \citet{KH2013}, there have been a number of studies which have provided new or updated estimates of BH mass in a variety of systems \citep[e.g.,][]{seth2014supermassive,saglia2016sinfoni,Krajnovic2018Supermassive,Thater2019Supermassive,nguyen2019DDBH,nguyen2019improved}. It is a high priority to obtain comparable bulge-disk decompositions in multiple bands and place these systems on similar scaling relations.


\subsection{Reverberation Mapped AGN}

The most accurate AGN BH masses are derived from reverberation mapping \citep[e.g.,][]{peterson2004central,2009bentz_reverberation,2011barth_reverberation}. In this technique the time lag between the continuum flux and broad line emission variability gives the light travel time across the Broad Line Region (BLR) of the AGN, and therefore gives the radius of the the BLR. With the dynamics of the BLR being dominated by the gravity of the central SMBH the radius and velocity of the BLR gas can be used to infer the mass of the central object. \par

In this work we include 37 galaxies hosting AGNs with reverberation mapped BH masses from \cite{bentz2018black}. \cite{bentz2018black} perform detailed morphological decomposition using GALFIT \citep{peng2010} with the goal of determining the bulge properties of their sample. They report stellar masses estimated using the $M/L$ ratios from \cite{belldejong2001} and \cite{into2013MLratio}. As the $M/L$ ratios derived in \cite{belldejong2001} and \cite{into2013MLratio} use differing IMFs than those derived by \citet{zibetti2009}, we cannot directly compare to the masses calculated in this work. Additionally, the colors used in the \cite{belldejong2001} and \cite{into2013MLratio} relations are different from those used in the \cite{zibetti2009} so we are unable to use the magnitudes reported by \cite{bentz2018black} to recalculate stellar masses. 

\begin{figure}
\centering
\includegraphics[width=\columnwidth]{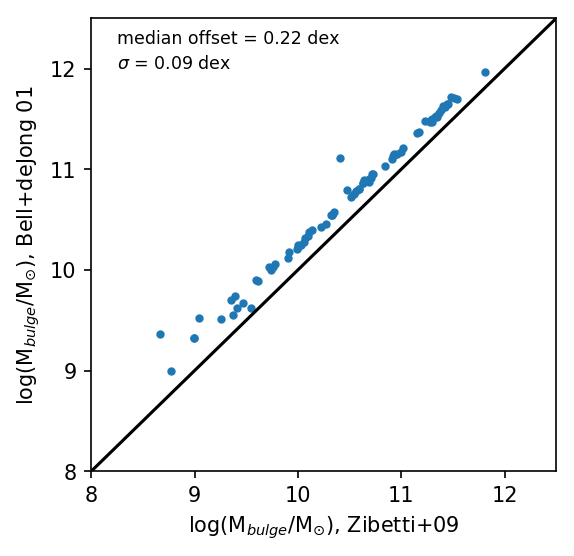}
\caption{Comparison of stellar masses of the sample from \cite{KH2013} derived from different methods. We estimate the mass-to-light ratios for K-band data as a function of B-V color following both \cite{zibetti2009} and \cite{belldejong2001}. The solid line show a one-to-one relation.}
\label{fig:KHmasscomp}
\end{figure}

To address this, we use the \cite{KH2013} sample to calculate stellar masses using the $M/L$ relation from \cite{belldejong2001} given by

\begin{equation}
\text{log} \left(M/L_K\right) = 0.652(B-V) - 0.692,
\end{equation}

\noindent
and then find the relation between these mass estimates and the stellar masses estimated using the \cite{zibetti2009} $M/L$ relations. We find that for the \citet{KH2013} sample the \cite{belldejong2001} relations predict stellar masses which have a median offset of 0.22 dex compared to the \cite{zibetti2009} estimates with scatter of $\sigma = 0.09$ dex. We then fit a linear regression to the different mass estimates, seen in Figure \ref{fig:KHmasscomp}, and find

\begin{equation}
    \text{log} \left(M_{\rm Zibetti}/M_{\odot}\right) = 1.06\text{log}\left(M_{\rm Bell+deJong}/M_{\odot}\right) - 0.91. 
\end{equation}

\noindent
We use this result to transform the bulge masses reported by \cite{bentz2018black} so they are consistent with stellar mass estimates found using the \citet{zibetti2009} relations and include them as part of our comparison sample.


\section{Black Hole Scaling Relations including Dwarf Galaxies}


We investigate BH-bulge (mass and luminosity) scaling relations using our sample of 7 dwarf galaxies with new {\it HST} observations, as well as the 125 additional galaxies considered in Section \ref{sec:additional_sys}.  The entire sample spans five orders of magnitude in BH mass and we have quadrupled the number of dwarf galaxies at the low-mass end.  For our sample of dwarf galaxies, we use the inner Sersic component as a proxy for the bulge without distinguishing between a classical bulge or pseudobulge.  We note that 12 dwarf galaxies in the full sample host broad-line AGNs and have BH masses estimated using single epoch spectroscopy, while the remaining objects have dynamical BH masses or reverberation-mapped AGN masses.  All bulge stellar masses are estimated in the most consistent way that is feasible, using color-dependent mass-to-light ratios from \citet{zibetti2009}. \par 


\subsection{Black Hole Mass - Bulge Stellar Mass Relation} \label{sec:MBHScaling}

\begin{figure*}
\centering
\includegraphics[width=0.7\textwidth]{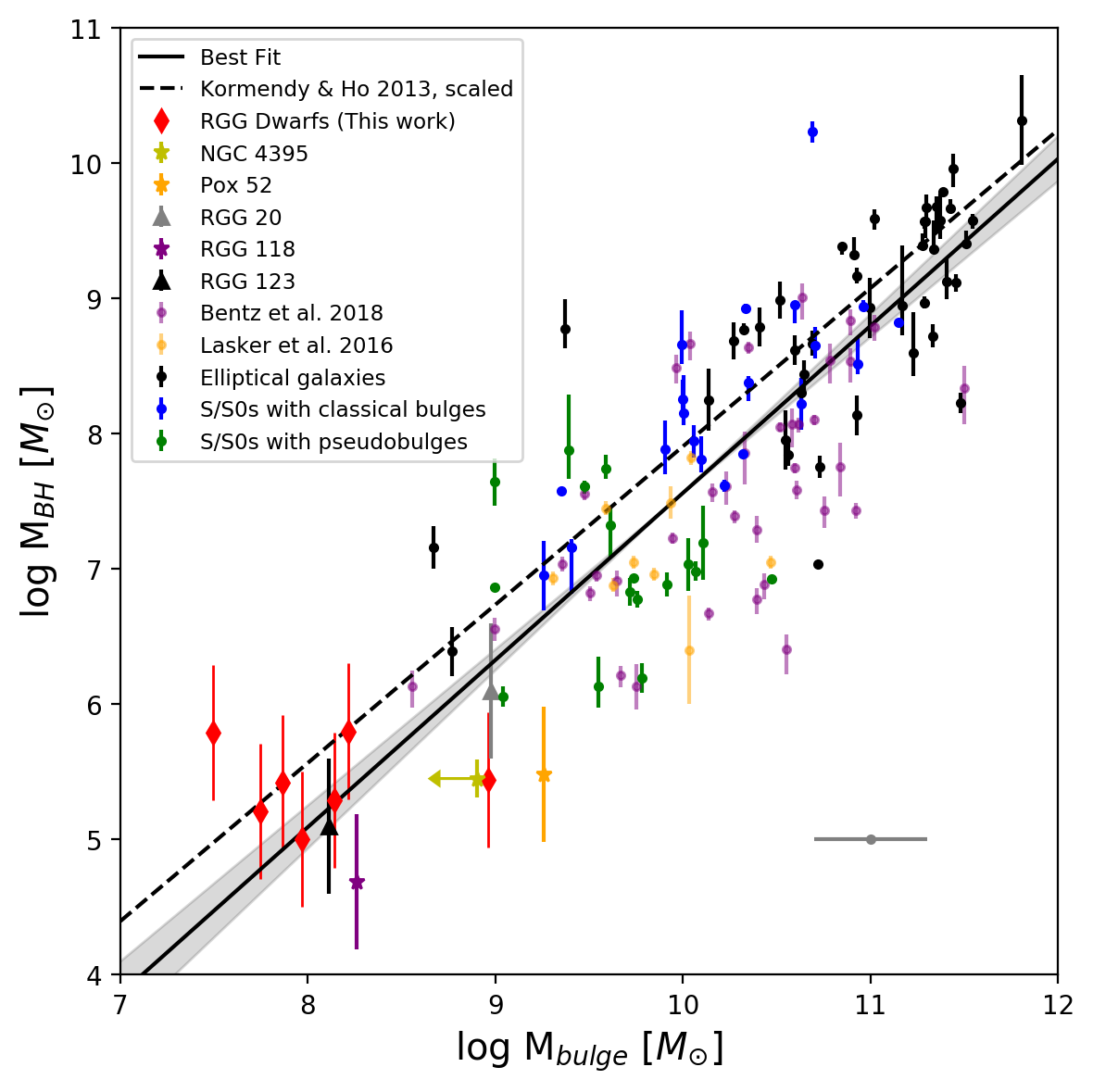}
\caption{Black hole mass versus bulge stellar mass. All bulge masses are estimated using color dependent mass-to-light ratios presented in \citet{zibetti2009}. Our sample of 7 broad-line AGN and composite dwarf galaxies from \citet{reines2013} with new {\it HST} observations are shown as red diamonds.  The dwarf galaxy RGG 118 \citep{reines2013,baldassare2015,baldassare2017} is shown as a pink star, POX 52 \citep{barth2004pox,thornton2008} is shown as an orange star and NGC 4395 \citep{filippenko1989} is shown as a green star with an arrow indicating it is an upper limit, it is not included in the linear regression. The dwarf galaxies RGG 20 and RGG 123 \citep[also see][]{greene2007activeBH,jiang2011host} are shown as purple and black triangles respectively. Dynamical BH mass measurements from \citet{KH2013} are shown as black (elliptical galaxies), blue (S/S0 galaxies with classical bulges) and green (S/S0 galaxies with pseudobulges) points. Late type, megamaser galaxies from \citet{lasker2016black} are shown as orange points. Galaxies with reverberation mapped AGNs from \cite{bentz2018black} are shown as purple points. The gray error bar indicates uncertainties in stellar mass estimates. The \citet{KH2013} relation was determined by fitting a sub-sample consisting of the elliptical and classical bulge systems reported in their work, whereas we employ their entire sample of 79 galaxies in the fitting performed for this work. The \citet{KH2013} ``scaled" relation has bulge masses scaled down by 0.33 dex to account for differences in adopted mass-to-light ratios (see \citealt{Reines2015}).}
\label{fig:scaling}
\end{figure*}


Figure \ref{fig:scaling} shows all 137 galaxies in our full sample on the $M_{\rm BH} - M_{\rm bulge}$ relation.  It is immediately clear that the dwarf galaxies in our study tend to fall on the extrapolation of the relation defined by more massive galaxies.  Whether this is expected or not is an open question, with some recent works finding evidence of low-mass or late-type systems falling below scaling relations derived using observations of high mass elliptical and classical bulge systems \citep{KH2013,lasker2016black}. We discuss our results in the context of a variety of observational and theoretical studies in \S\ref{sec:ScalingDiscuss}. \par

We find that a linear relationship between log($M_{\rm BH}$) and log($M_{\rm bulge}$) is a good description of the data shown in Figure \ref{fig:scaling}, with a Spearman correlation coefficient $\rho=0.81$ and a probability $p < 10^{-31}$ that no linear correlation is present. We therefore parameterize the scaling relation as

\begin{equation}
\text{log}(M_{\rm BH}/M_{\odot}) = \alpha + \beta \text{log}(M_{\rm bulge,\star}/10^{11}M_{\odot})
\label{eqn:scale}
\end{equation}

\noindent
for direct comparison to other studies. We perform a linear regression using the Bayesian approach implemented in the LINMIXERR algorithm developed by \cite{kelly2007linmixerr}. This process allows for the inclusion of measurement errors in both variables, as well as accounting for a component of intrinsic, random scatter. We report values for the slope, intercept and scatter for this relationship which are the median values and 1$\sigma$ widths of a large number of draws from the posterior distribution for each quantity. Using this method, we find best fit parameters of

\begin{equation}
\alpha = 8.80 \pm 0.085; \beta = 1.24 \pm 0.081,
\label{eqn:coef}
\end{equation}

\noindent
where the scatter about the relation has a standard deviation of $\sigma=0.68$ dex. 

To illustrate the impact of using different $M/L$ ratios, we refit the relation using bulge stellar masses derived using the color dependent mass-to-light ratios from \citet{bell2003} (using the same relations discussed in \S\ref{sec:stellarmasses} and \S\ref{sec:additional_sys} with coefficients found in Table A7 from \citet{bell2003}). We find a best fit of 

\begin{equation}
\alpha = 8.53 \pm 0.076; \beta = 1.41 \pm 0.094
\label{eqn:coef2}
\end{equation}

\noindent
where the scatter about the relation has a standard deviation of $\sigma=0.70$ dex. 


\subsection{Black Hole Mass - Bulge Luminosity Relation}\label{sec:LBHScaling}

\begin{figure}
\centering
\includegraphics[width=\columnwidth]{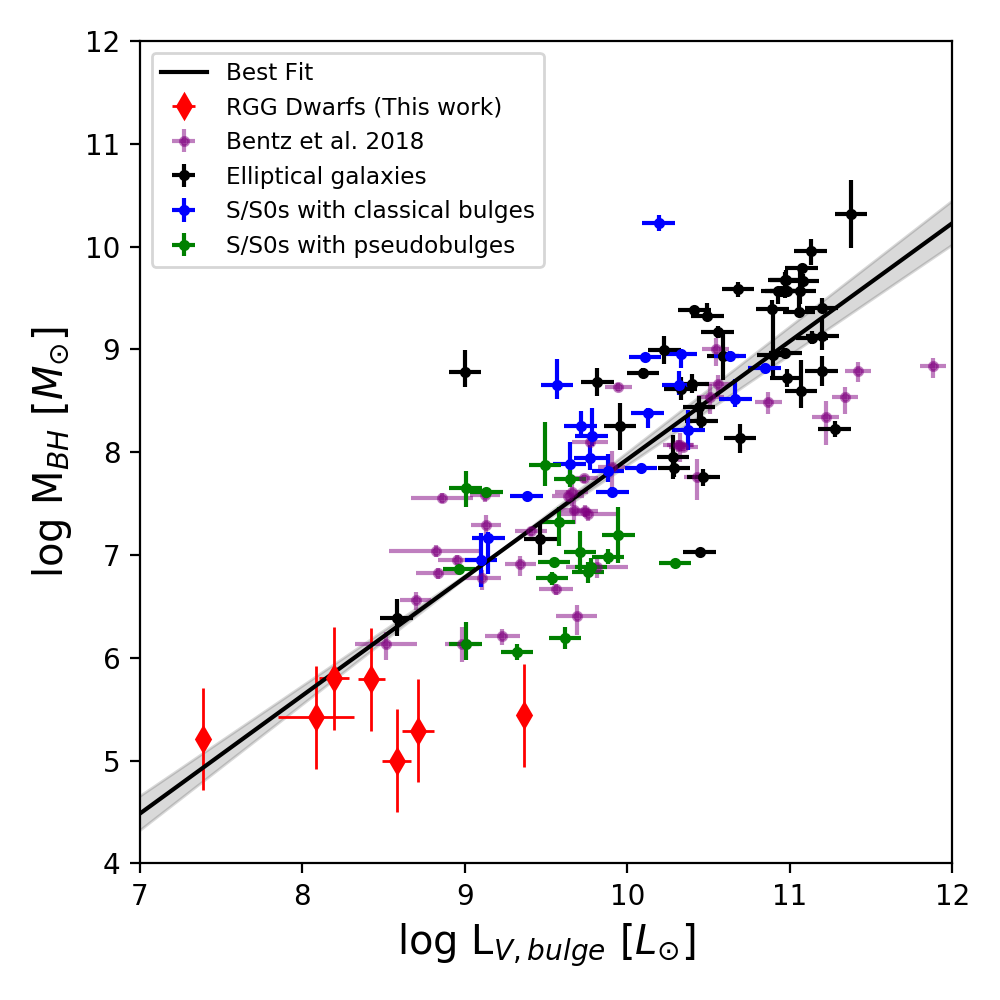}
\caption{Black hole mass versus bulge V band luminosity. Our sample of 7 broad-line AGN and composite dwarf galaxies from \citet{reines2013} with new {\it HST} observations are shown as red diamonds. Dynamical BH mass measurements from \citet{KH2013} are shown as black (elliptical galaxies), blue (S/S0 galaxies with classical bulges) and green (S/S0 galaxies with pseudobulges) points. Galaxies with reverberation mapped AGNs from \cite{bentz2018black} are shown as purple points. The best fit linear regression to our sample of dwarf galaxy systems and the entire comparison sample is shown as the solid black line, with the gray shading corresponding to 1 $\sigma$ uncertainties in the fit parameters.}
\label{fig:MBHLBulge}
\end{figure}

One of the earliest BH scaling relations to be discovered was the BH mass and bulge luminosity, $M_{\rm BH}-L_{\rm bulge}$, relation \citep{KormendyRichstone1995,Magorrian1998Scaling}. While this relationship is thought to arise as a consequence of the $M_{\rm BH}-\sigma$ and $M_{\rm BH}-M_{\rm bulge}$ relations \citep{KH2013}, investigating this relation remains a relevant concern. Recent work has led to much tighter $M_{\rm BH}-L_{\rm bulge}$ relations with reported scatter about the relation becoming similar to that found in the $M_{\rm BH}-\sigma$ relation \citep{marconi2003relation,gultekin2009scaling}. Other work has also expanded the scope of this relation to include not only bulge-dominated classical and elliptical systems, but also late-type galaxies \citep{wandel2002black,bentz2009_MBHLscaling,bentz2018black}. Given the variety of methods used to estimate bulge stellar mass and the accompanying uncertainties, the $M_{\rm BH}-L_{\rm bulge}$ provides useful information and additional constraints on BH-bulge scaling relations. \par

To investigate this relation in the context of our sample of dwarf galaxies, we estimate the $V$-band luminosity of the inner Sersic component for our sample using the same photometric conversion technique described in \S\ref{sec:PhotConversion}. The $V$-band luminosities are used as this allows us to use the absolute $V$-band magnitudes reported by \cite{KH2013} and the $V$-band luminosities reported in \cite{bentz2018black} to act as comprehensive comparison sample.

We find strong evidence of a linear correlation between the BH mass and the $V$-band luminosity, with a Spearman correlation coefficient $\rho=0.83$ and a probability $p < 10^{-32}$ that no correlation is present. With this consideration we parameterize the scaling relation as

\begin{equation}
\text{log}(M_{\rm BH}/M_{\odot}) = \alpha + \beta \text{log}(L_{V,\rm bulge}/10^{10}L_{\odot})
\label{eqn:scale2}
\end{equation}

\noindent and find a best fit of

\begin{equation}
\alpha = 7.93 \pm 0.061; \beta = 1.15 \pm 0.075
\label{eqn:coef3}
\end{equation}

\noindent where the scatter about the relation has a standard deviation of $\sigma = 0.67$. 
This result, seen in Figure \ref{fig:MBHLBulge}, is in good agreement with previous works that examine the relation between BH mass and optical bulge luminosity. \cite{marconi2003relation} use a sample of 27 galaxies to investigate this scaling relation for near IR and optical spectral bands. When considering the optical $B$-band they find the slope of their relation to be $\beta = 1.19 \pm 0.12$ with an intercept of $\alpha = 8.18 \pm 0.08$. \cite{McconnellMaScaling2013} study 72 BHs and their host galaxies to study a variety of scaling relations. When considering the relation between BH mass and bulge $V$ band luminosity they find a slope of $\beta = 1.11 \pm 0.13$ and an intercept of $\alpha = 8.12 \pm 0.10$. Recently \cite{bentz2018black} study this relation using the bulge $V$-band luminosities to from their sample and those from \cite{KH2013} to find a slope of $\beta = 1.13 \pm 0.08$ and an intercept of $\alpha = 8.04 \pm 0.06$. 

\section{Discussion}\label{sec:ScalingDiscuss}

While we find evidence of a power law $M_{\rm BH} - M_{\rm bulge}$ relation holding to into the low-mass regime, there has been some evidence of the BHs in spiral and dwarf galaxies falling below the scaling relations derived using samples of elliptical and classical bulge galaxies. Recently this trend was observed by \citet{lasker2016black} when studying the host galaxies of megamasers, finding that the BHs in their sample fall low with respect to both BH-bulge mass and BH-total mass relations. The results presented here seem to be at odds with this result, as we find evidence for the BH-bulge mass relation holding to BH masses of $\sim 10^5 M_{\odot}$. Additionally, recent work done by \citet{nguyen2019improved} studying three nearby early type dwarf galaxies with BHs masses estimated via dynamical modeling find evidence that these systems also fall below power law scaling relations derived when considering higher mass systems. While it is not evident why this disagreement occurs there are several possible explanations for these discrepancies. \par

One possibility is that the megamaser galaxies studied by \cite{lasker2016black} are biased towards a lower $M_{\rm BH}$ at a fixed galaxy property, as the megamaser disk may select galaxies which are in the process of growing toward the end state of these scaling relations. \citet{lasker2016black} argue against this, citing \citet{greene2010scaling} who point out that in order for the BHs in the megamaser galaxies to grow enough to move onto observed scaling relations they would have to accrete at $\sim10\%$ Eddington for 1 Gyr, longer than the expected lifetime of an AGN \citep{martini2001quasar,greene2016megamaser}. \par

A more plausible explanation for the differences in our findings could be that there is a bias towards finding higher mass BHs in low mass galaxies. This bias could manifest from an observational standpoint as it will be much easier to detect the highest mass systems from an underlying distribution of BHs. One aspect of this is discussed by \citet{reines2013} who find that to detect broad H$\alpha$ emission with a detection limit of $\sim 10^{-15}$erg s$^{-1}$ cm$^{-2}$, a BH of $\sim 8 \times 10^3 M_{\odot}$ would have to be accreting at its Eddington limit. Since consistent Eddington limited accretion is unlikely, there is an observational bias towards finding more massive BHs accreting at modest rates. \par 

The final source of this discrepancy we consider is possible errors in the BH mass estimations for our sample of dwarf galaxies. The BH masses estimated with virial techniques using single-epoch virial methods are very indirect (see \S\ref{sec:BHmasses}) and subject to many systematic uncertainties. An obvious case of this is that the BLR geometry will vary from object to object \citep{kollatschny2003accretion,2009bentz_reverberation,denney2010reverberation,2011barth_reverberation} which makes the use of a single geometric scaling factor suspect. While the systematic uncertainties in the BH estimates for our sample may resolve the observed differences to the megamaser sample studied by \citet{lasker2016black} there is still the discrepancy with trends seen by \cite{nguyen2019improved}, who estimate BH masses using dynamical modeling, which is subject to fewer systematic uncertainties. Clearly to resolve this issue, improved observations of galaxies hosting intermediate mass BHs are needed. \par

While there are discrepancies between our findings and some recent work, our derived scaling relation between BH mass and bulge mass is in reasonable agreement with a variety of other studies (see Figure \ref{fig:scaling}). \cite{haringrixscaling2004} investigate this scaling relation using a sample of 30 galaxies, finding a slope of $\beta = 1.12 \pm 0.06$. Similarly, when considering elliptical and classical bulge systems, \cite{KH2013} find the slope of this relation to $\beta = 1.16 \pm 0.08$. \cite{McconnellMaScaling2013} find a similar range of slopes from $\beta = 1.05 \pm 0.11$ to $\beta = 1.23 \pm 0.16$ depending on how the stellar mass is estimated (dynamics versus stellar populations). \cite{saglia2016sinfoni} investigate a number of BH-host galaxy scaling relations using a database of 97 galaxies which contains a variety of galaxy morphologies. When considering the BH mass-bulge mass scaling relation for the entire sample they find a slope of $\beta = 0.96 \pm 0.07$. Most recently, \cite{bentz2018black} used a sample of 37 reverberation mapped AGN plus galaxies from \cite{KH2013} and nine megamaser galaxies from \cite{lasker2016black} to find a slope of $\beta = 1.50 \pm 0.13$. \par


\begin{figure}
\centering
\includegraphics[width=\columnwidth]{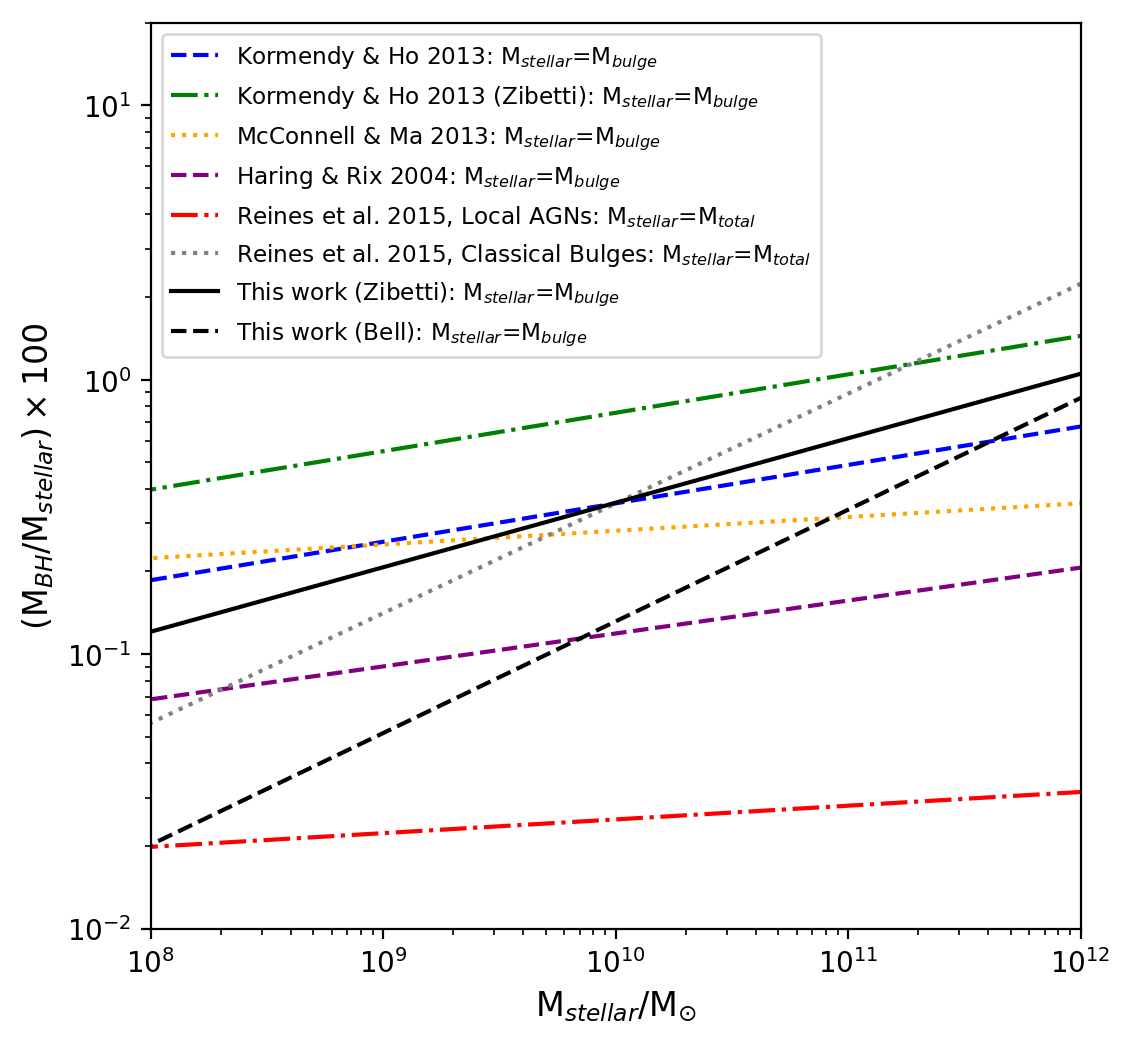}
\caption{BH mass fractions (given as a percentage of the bulge/stellar mass) as a function of bulge/stellar mass. Our relation including RGG dwarf galaxies is shown as a solid black line. Bulge and stellar mass relations from literature are shown for comparison.}
\label{fig:Offsetcomp}
\end{figure}


In addition to considering the slope of these relations, it is important to compare the intercepts found as well. Offsets between intercepts will result in large discrepancies in BH mass estimates for relations with similar slopes. This is exemplified in Figure \ref{fig:Offsetcomp}, where we compare a re-scaling of the BH-bulge mass relation which gives the BH to bulge (or stellar) mass ratio as a function of bulge (or stellar) mass from several different studies. With the relation parameterized in this way it is evident that small changes in the intercept of a scaling relation can result in large changes in the estimated BH to bulge mass ratio. We find that our intercept is in good agreement with many studies which consider the BH-bulge mass relation. \cite{haringrixscaling2004} find an intercept of $\alpha = 8.20 \pm 0.10$ when considering the sample of 30 galaxies mentioned above. When investigating a sample of megamaser galaxies \cite{lasker2016black} find an intercept of $\alpha = 8.12 \pm 0.08$ for their most detailed morphological decomposition. Scaling the relation derived by \cite{saglia2016sinfoni} using their entire sample to match the convention used in this paper results in an intercept of $\alpha = 8.48 \pm 0.716$. Finally, when the relation from \cite{bentz2018black} is re-scaled to match the convention used here they find an intercept of $\alpha = 8.66 \pm 0.11$. \par

We also compare our results to cosmological simulations, noting that there are a variety of methods used to estimate bulge masses in these works \citep[e.g.,][]{degraf2015scaling,SijackiIllustris2015,SchayeEagle2015}.
\cite{SijackiIllustris2015} investigated the $M_{\rm BH}-M_{\rm bulge}$ scaling relation using results from the high-resolution Illustris simulations. They estimate the the bulge stellar mass of galaxies in their simulation to be the stellar mass within the stellar half-mass-radius. This does not take into account the varying bulge-mass-fractions of galaxies, in addition to neglecting cases of bulgeless or elliptical galaxies. While the proxy for bulge stellar mass may not be entirely robust, the Illustris simulation assumes a Chabrier IMF, which is the same IMF assumed by \cite{zibetti2009} and aids in a more direct comparison with our results.  Within their simulation sample, \cite{SijackiIllustris2015} report a slope of $\beta=1.21$  and when the relation is re-scaled to match our normalization an intercept of $\alpha = 8.02$ which is in good agreement with our results. \par

The  agreement  found  between  a  variety  of  observational samples suggests a robust scaling relationship be-tween BH mass and bulge mass.  The scatter among relationships  seen  in  Figure  5  seems  to  be  driven  by  use of different stellar mass estimators and differing definitions of stellar mass (bulge mass vs.\ total stellar mass). Both this work and \citet{Reines2015} find that contributions from AGN light have only marginal effects on the estimation of stellar mass, at least for moderate luminosity  Seyferts. This is in contrast to the differences in stellar mass estimates when using various color dependent mass-to-light ratios (i.e.\ relations from 
\citet{bell2003} or \citet{zibetti2009}) or the large difference we find between bulge mass and total stellar mass. With these considerations in mind it is worth emphasizing that one must be careful when selecting ‘bulge’ mass, which can vary substantially from the total stellar mass of a galaxy. This is readily apparent in the sample  of seven dwarf galaxies studied in this paper, for which we find a median bulge-to-total stellar mass ratio of only ∼ 0.1. Assuming total stellar mass is a good proxy for bulge mass, particularly at the low-mass end, can substantially impact the derived scaling relation and subsequent inferences. For example, \citet{graham2014black} use total stellar masses for the 10 broad-line AGN and composite dwarf galaxies in \citet{reines2013}, all of which are included in this work. Under the assumption that total stellar mass is equivalent to bulge stellar mass, they conclude that AGN host galaxies follow an approximately quadratic relation between BH mass and bulge stellar mass.  This is in disagreement with our study and highlights the important distinction between bulge stellar mass and total stellar mass in dwarf galaxies, as well as the need for high resolution imaging to properly perform structure decomposition for these systems.


\section{Summary and Conclusions} \label{sec:conclusions}

We have presented high-resolution optical and near IR \textit{HST} images of 7 dwarf galaxies hosting broad-line AGNs with single-epoch spectroscopic BH mass estimates. We find that 6 of the 7 active dwarf galaxies in our sample have a photometric structure consistent with a bulge/pseudobulge and exponential disk, with the disk dominating the stellar mass of the system. This photometric decomposition allows us to compare our sample to more massive systems with dynamically determined BH masses \citep{KH2013,lasker2016black} and reverberation-mapped AGNs \citep{bentz2018black}, allowing for a re-examination of the $M_{\rm BH}-M_{\rm bulge}$ and $M_{\rm BH}-L_{\rm bulge}$ relations with BH masses which are an order of magnitude lower than in previous studies. With the inclusion of low-mass systems and active galaxies, this work offers robust estimates for estimating BH masses over a broad range of galaxy properties. \par

Overall we find the inclusion of our dwarf galaxy sample results in a $M_{\rm BH} - M_{\rm bulge}$ relation which is linear and has a slope of  $\beta = 1.24 \pm 0.09$ and an intercept of $\alpha = 8.80 \pm 0.09$, which is in good agreement with previous relations based on more massive quiescent and active galaxies \citep{haringrixscaling2004,KH2013,McconnellMaScaling2013,bentz2018black}, as well as the results from cosmological hydrodynamical simulations \citep[e.g.,][]{SijackiIllustris2015}. On the other hand, our results are in conflict with some recent studies finding low-mass and late-type galaxies falling below the BH-bulge mass scaling relation \citep{lasker2016black,nguyen2019DDBH}.  Given the observational bias towards finding more massive BHs in dwarf galaxies, it is plausible that additional low-mass systems will be found to fall below the relation with more sensitive searches.  

This work has quadrupled the number of active dwarf galaxies 
on the BH-bulge mass relation, and we are reaching the mass regime where the signatures of BH seeds are expected to manifest. 
Modeling of various BH seeding scenarios \citep{volonteri2008evolution,greene2012low,natarajan2014seeds} have found evidence that if BH seeds are heavy ($M_{\rm BH,seed}\approx 10^{4-5}~M_{\odot}$, as is predicted from `direct collapse' models) the low mass end of scaling relations between BH mass and host galaxy properties will flatten creating a `plume' of ungrown BHs. Alternately, if BH seeds are light ($M_{\rm BH,seed}\approx 100~M_{\odot}$, as is predicted by models of the collapse of Population III stars) the characteristic `plume' of ungrown BHs would scatter below observed scaling relations. While the scatter about the $M_{\rm BH}-M_{\rm bulge}$ scaling relation presented here is somewhat larger than that when only considering more massive elliptical and classical bulge systems \citep[e.g.,][]{KH2013}, we do not observe a distinct `plume' above or below the relation. These considerations highlight the need to search for even lower mass systems (such as RGG 118 studied in depth by \citealt{baldassare2017}) to further constrain the formation of the first massive BHs.

\vspace{10mm}

We thank Vivienne Baldassare and Chien Peng for input on the use of GALFIT and modeling our galaxy images. We also thank Marla Geha, Julie Comerford and Laura Blecha for helpful discussions. Additionally we thank the anonymous referee for thoughtful and helpful comments. Support for program number HST-GO-13943.007-A was provided by NASA through a grant from the Space Telescope Science Institute, which is operated by the Association of Universities for Research in Astronomy, Incorporated, under NASA contract NAS5-26555. Based on observations made with the NASA/ESA Hubble Space Telescope, obtained at the Space Telescope Science Institute, which is operated by the Association of Universities for Research in Astronomy, Inc., under NASA contract NAS5-26555. These observations are associated with program number HST-GO-13943.007-A.


\appendix

\section{Notes on individual galaxies} \label{sec:gal_decomp}

\subsection{RGG 1} \label{app:RGG1}

RGG 1 (Figure \ref{fig:RGG1_Model}) is a S0 galaxy that contains a disky outer component ($n_{\rm outer} \sim 0.8$) with a half-light radius of $\sim$1.6 kpc. The inner Sersic component has a half-light radius of $\sim$0.7 kpc with a roughly Gaussian profile ($n_{\rm inner} \sim 0.3$). The disk component has an axis ratio of 0.7 and the inner component has an axis ratio of 0.5. The inner Sersic component is subdominant to the disk at all radii, which may indicate a nuclear disk as opposed to a more classical bulge component. With these factors in mind we classify RGG 1 as pseudobulge galaxy. The point source included to model the AGN has a low central surface brightness which is consistent with the X-ray observations from \citet{baldassare2017x} who find the Eddington ratio to be $L_{Bol}/L_{Edd} \sim 0.001$.

\subsection{RGG 9} \label{app:RGG9}

\begin{figure*}
\centering
\subfloat{\includegraphics[width=\textwidth]{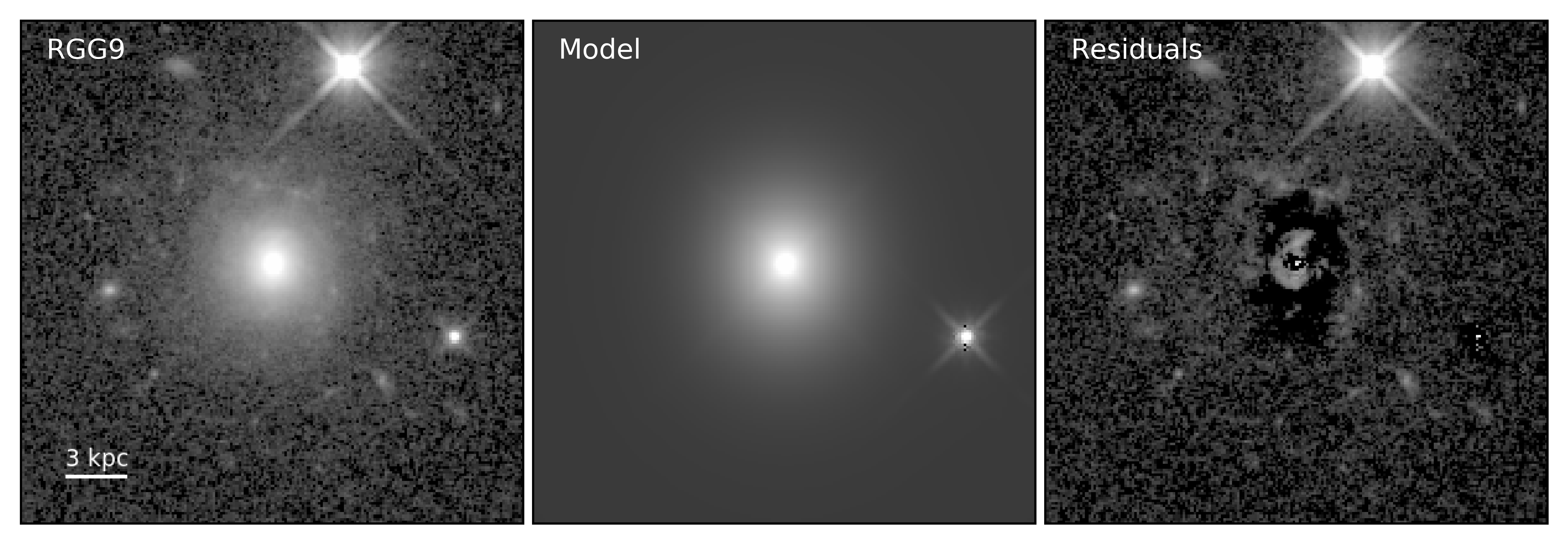}}
\vspace{-0.1cm}
\begin{minipage}{.47\linewidth}
\centering
\subfloat{\includegraphics[scale=0.47]{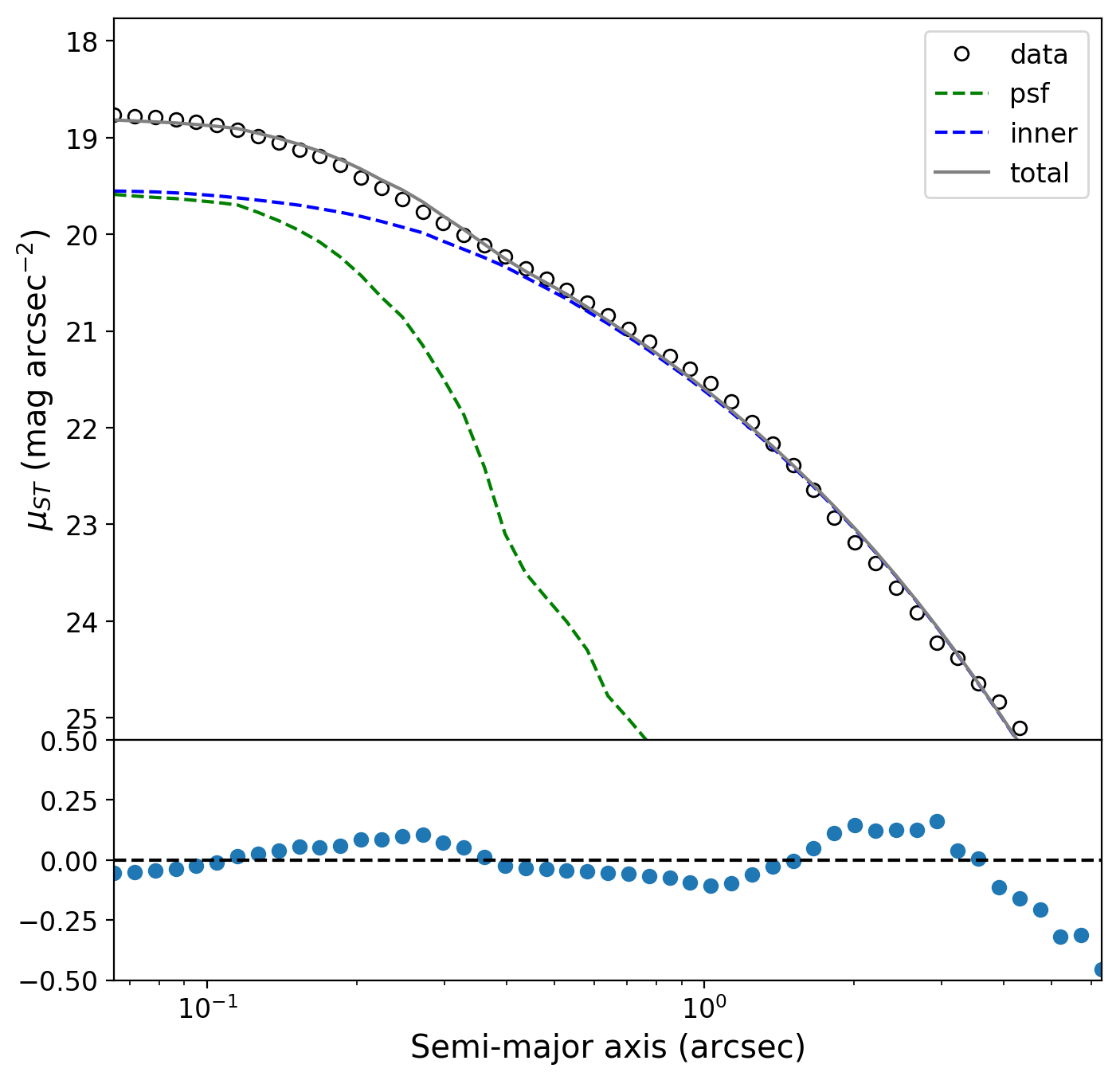}}
\end{minipage}
\hfill
\begin{minipage}{.47\linewidth}
\centering
\subfloat{\includegraphics[scale=0.4]{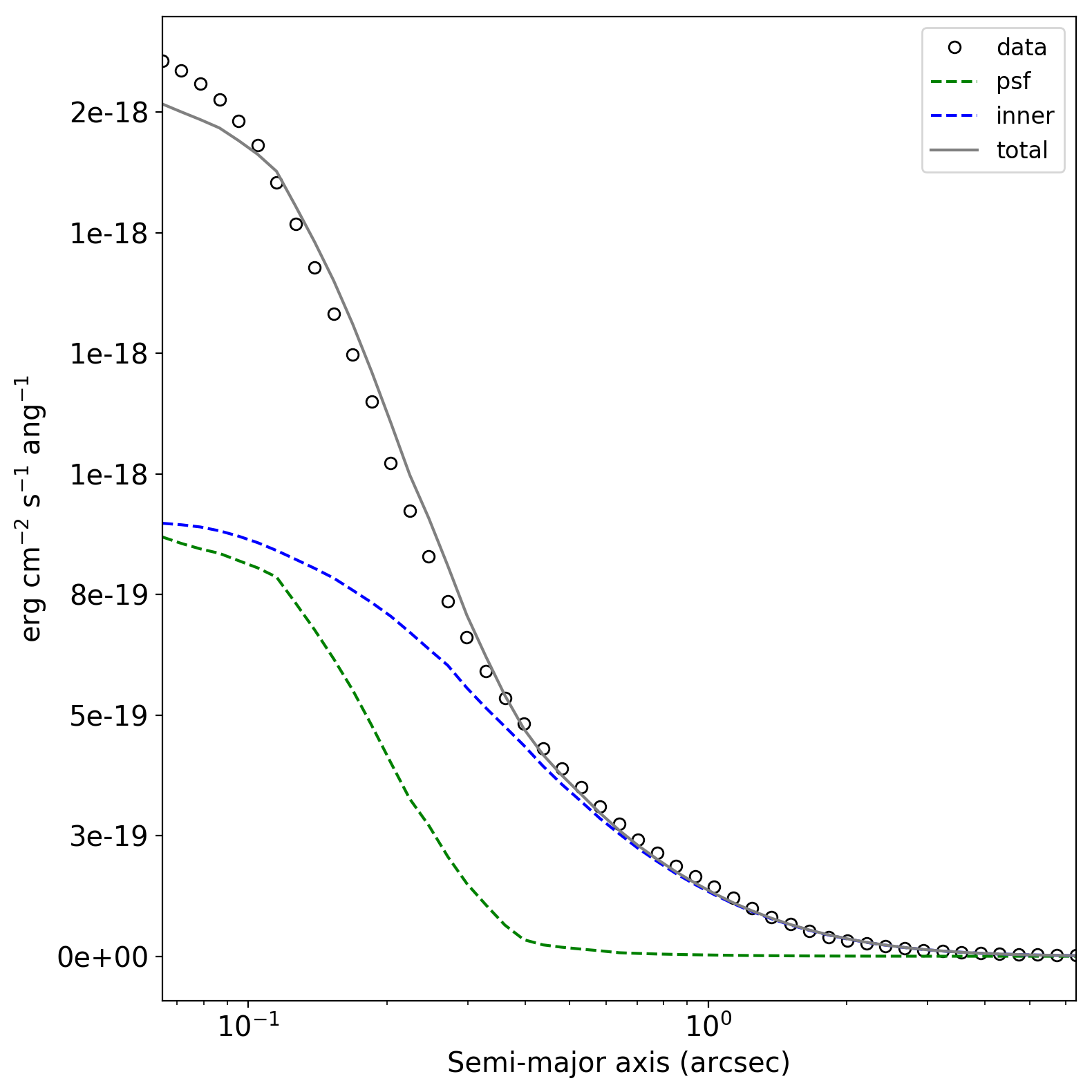}}
\end{minipage}
\caption{Top row: image of RGG 9 in the F110W filter (left); best fit GALFIT model which includes a PSF, inner Sersic component and outer Sersic component (middle); residuals (right). Bottom row: Left panel shows the observed surface brightness profile of RGG 9 with open circles. The best fit model is shown in gray, with the components being shown in green (PSF) and blue (inner Sersic). The residuals are shown in the lower panel. The right panel shows the average intensity along a given isophote for the data and the intensity as a function of radius. }
\label{fig:RGG9_Model}
\end{figure*}

RGG 9 (Figure \ref{fig:RGG9_Model}) appears to be a dwarf elliptical galaxy with a nuclear disk. The nuclear disk is most notable when examining the residuals seen in the third panel of Figure \ref{fig:RGG9_Model}. Though the disk is clearly visible in the residuals in the F110W filter this feature is more difficult to fit in the F606W filter, and including a component to account for the disk produces a mass for the nuclear disk which is overly large ($M_{ND} \approx 10^{9}M_{\odot}$). When we choose to fit a single Sersic component we find the magnitudes do not change significantly ($\Delta m_{ST}^{elliptical} \approx 0.05$) in either filter. Similarly the Sersic index of the elliptical component does not change drastically either, from $n\approx 4$ to $n\approx 2.5$ when going from the nuclear disk/elliptical fit to a single Sersic component. With this in mind we use a single Sersic component fit and find the elliptical portion to have a half-light radius of $\sim$1.21 kpc with and axis ratio of $\sim$0.85. As this is a dwarf elliptical galaxy it is classified as a classical bulge.

\subsection{RGG 11} \label{app:RGG11}

\begin{figure*}
\centering
\subfloat{\includegraphics[width=\textwidth]{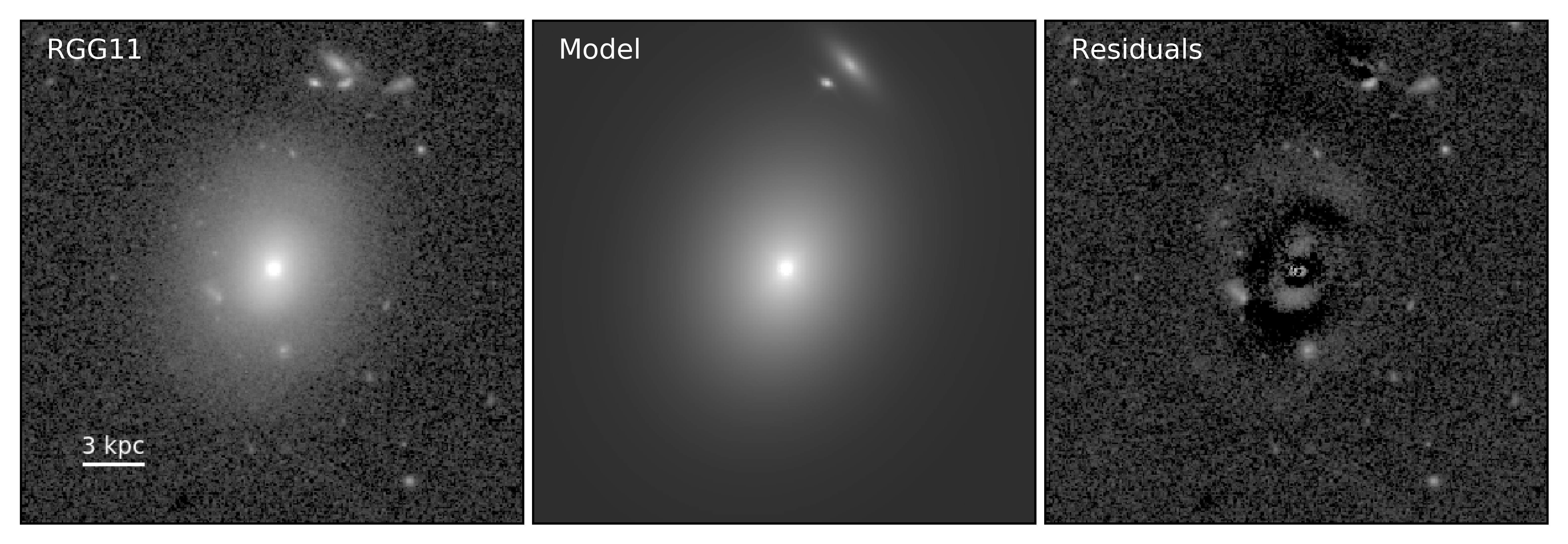}}
\vspace{-0.1cm}
\begin{minipage}{.47\linewidth}
\centering
\subfloat{\includegraphics[scale=0.47]{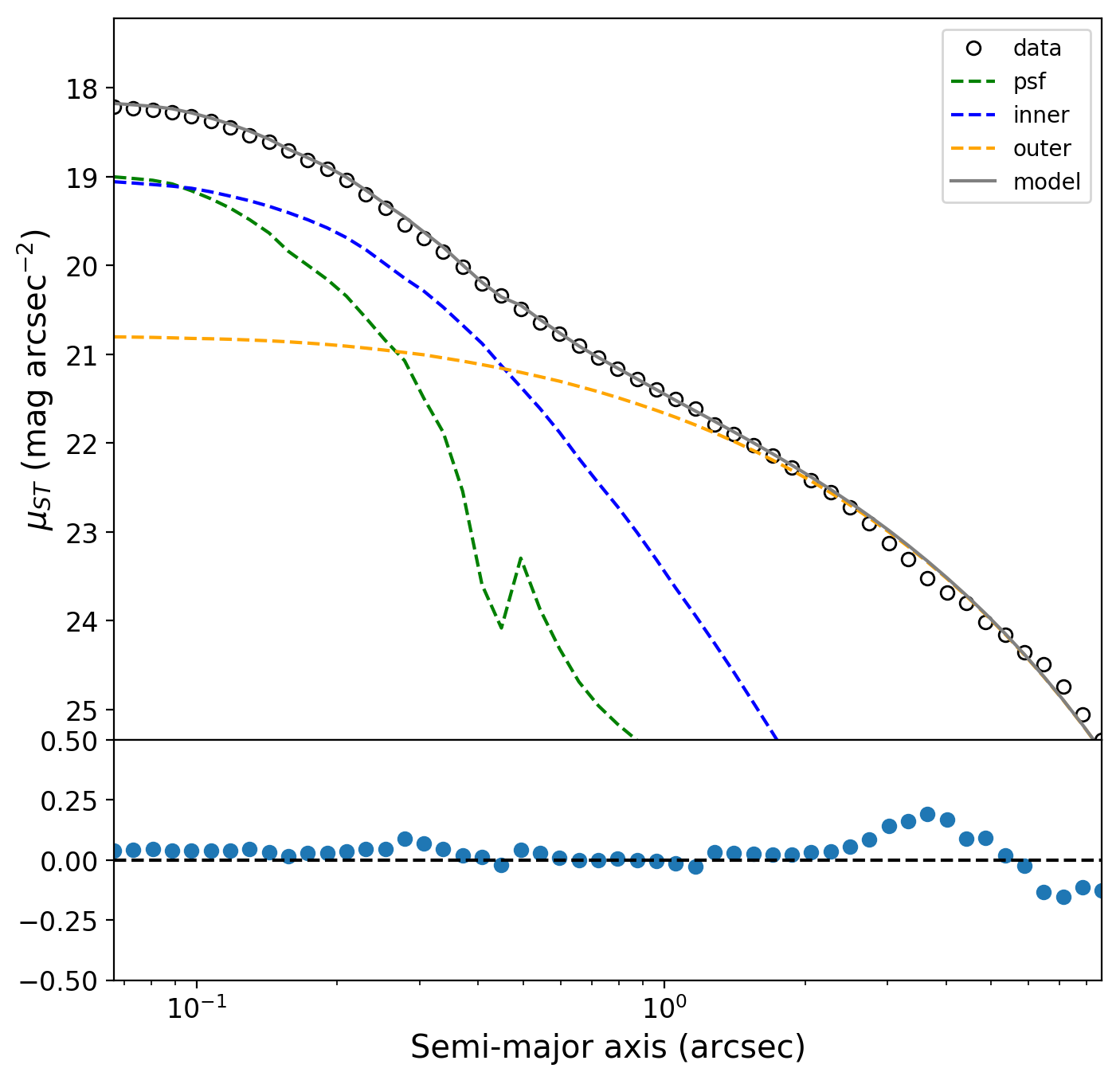}}
\end{minipage}
\hfill
\begin{minipage}{.47\linewidth}
\centering
\subfloat{\includegraphics[scale=0.4]{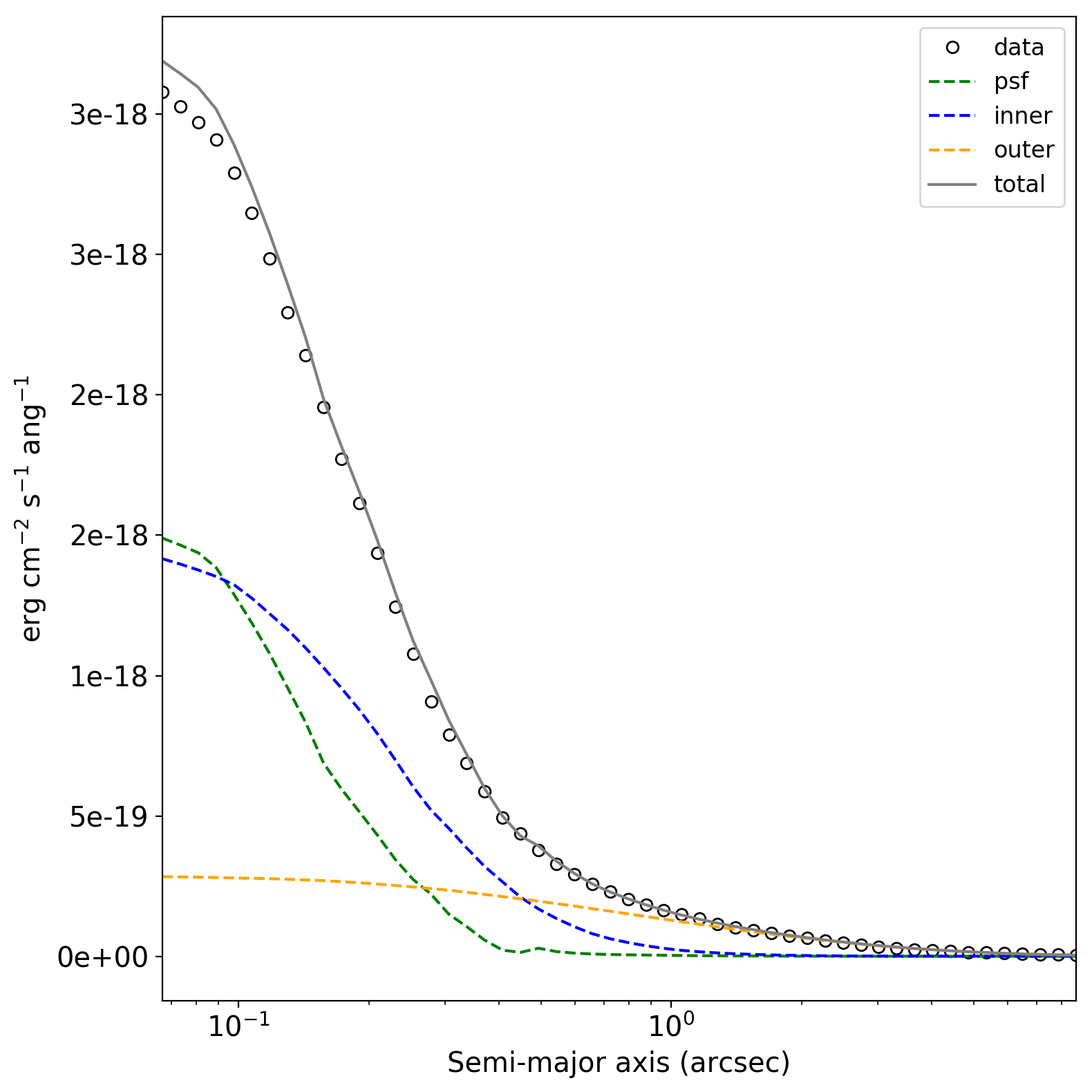}}
\end{minipage}
\caption{Top row: image of RGG 11 in the F110W filter (left); best fit GALFIT model which includes a PSF, inner Sersic component and outer Sersic component (middle); residuals (right). Bottom row: Left panel shows the observed surface brightness profile of RGG 11 with open circles. The best fit model is shown in gray, with the components being shown in green (PSF), blue (inner Sersic) and orange (outer Sersic). The residuals are shown in the lower panel. The right panel shows the average intensity along a given isophote for the data and the intensity as a function of radius. }
\label{fig:RGG11_Model}
\end{figure*}

RGG 11 (Figure \ref{fig:RGG11_Model}) appears to be an S0 galaxy with galaxy with classical structure. The inner component is rounded and has a Sersic index of $n = 2.4$, indicating a classical bulge. It has a half-light radius of 0.13 kpc. The outer component has a slightly higher than average Sersic index with $n=1.69$, but still closely resembles the classic exponential disk. The disk is one of the larger features in our sample of galaxies with a half-light radius of 2.57 kpc.

\subsection{RGG 32} \label{app:RGG32}

\begin{figure*}
\centering
\subfloat{\includegraphics[width=\textwidth]{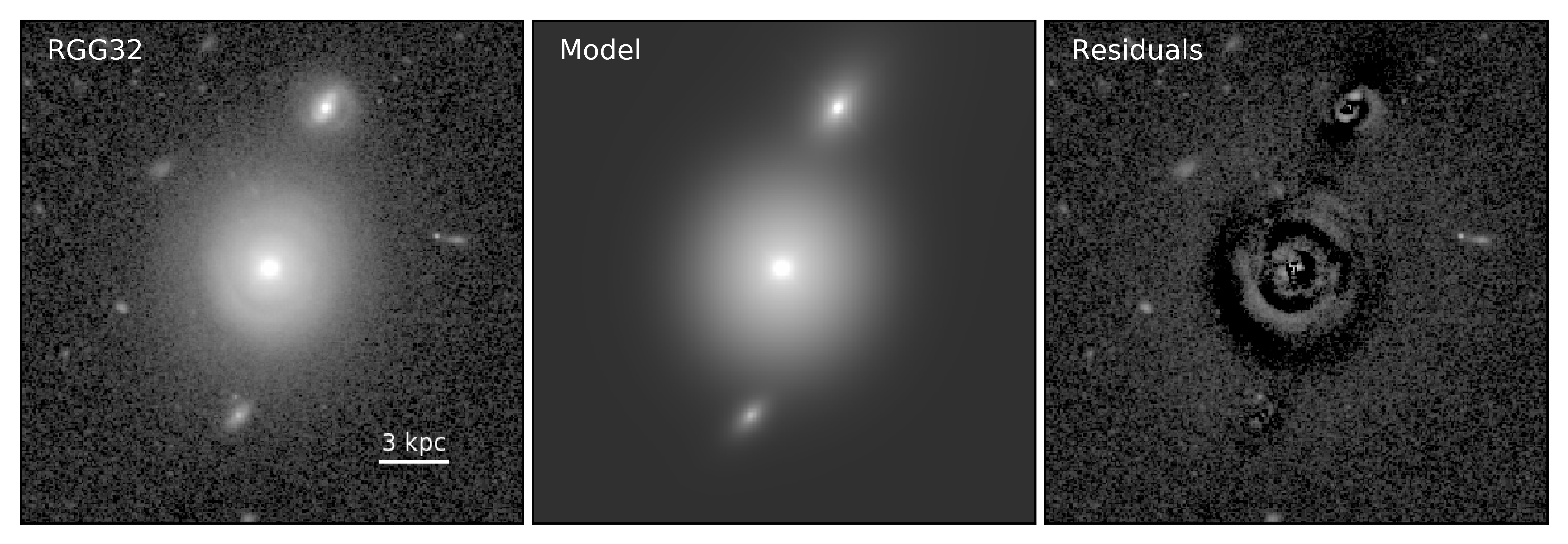}}
\vspace{-0.1cm}
\begin{minipage}{.47\linewidth}
\centering
\subfloat{\includegraphics[scale=0.47]{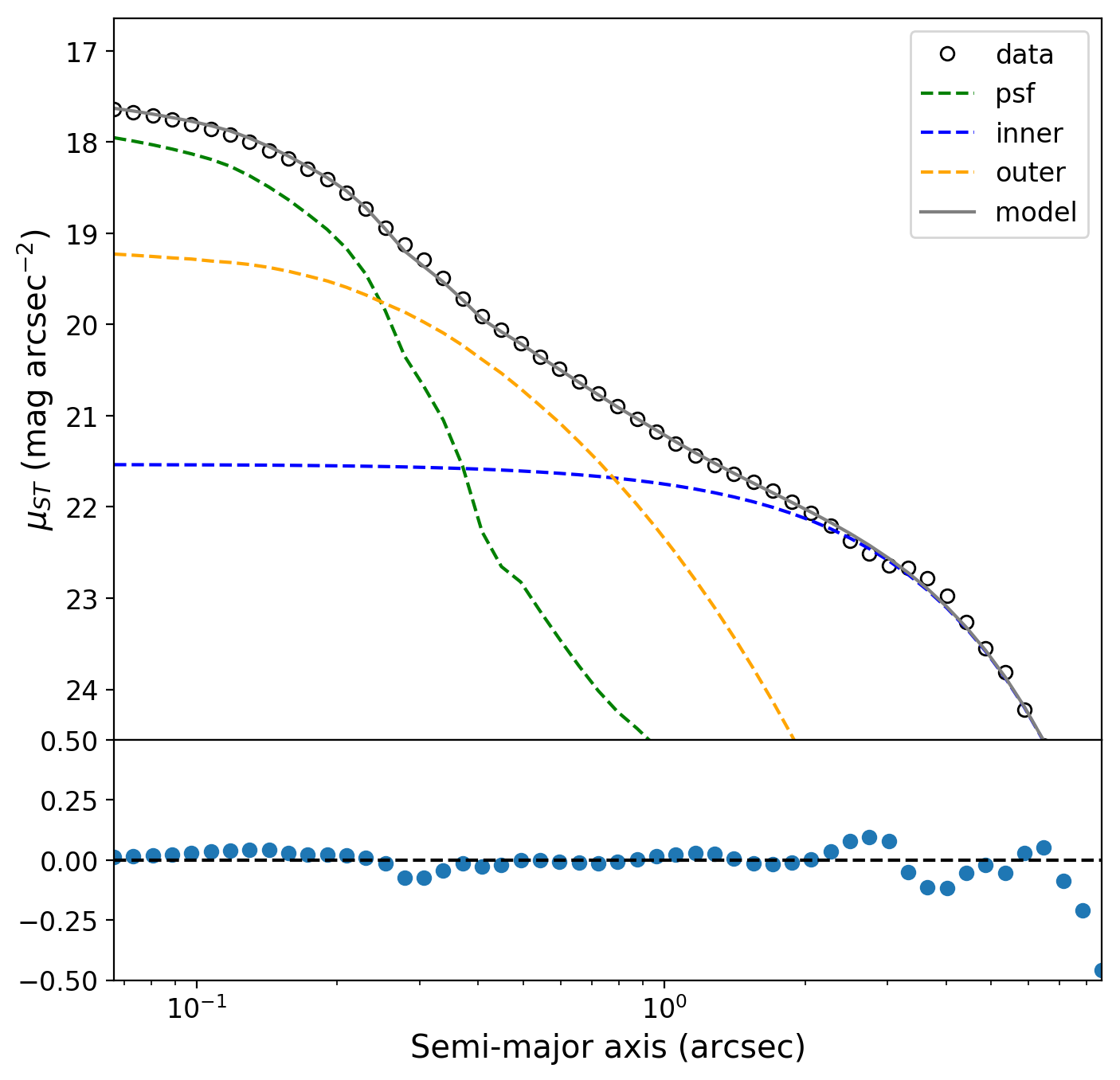}}
\end{minipage}
\hfill
\begin{minipage}{.47\linewidth}
\centering
\subfloat{\includegraphics[scale=0.4]{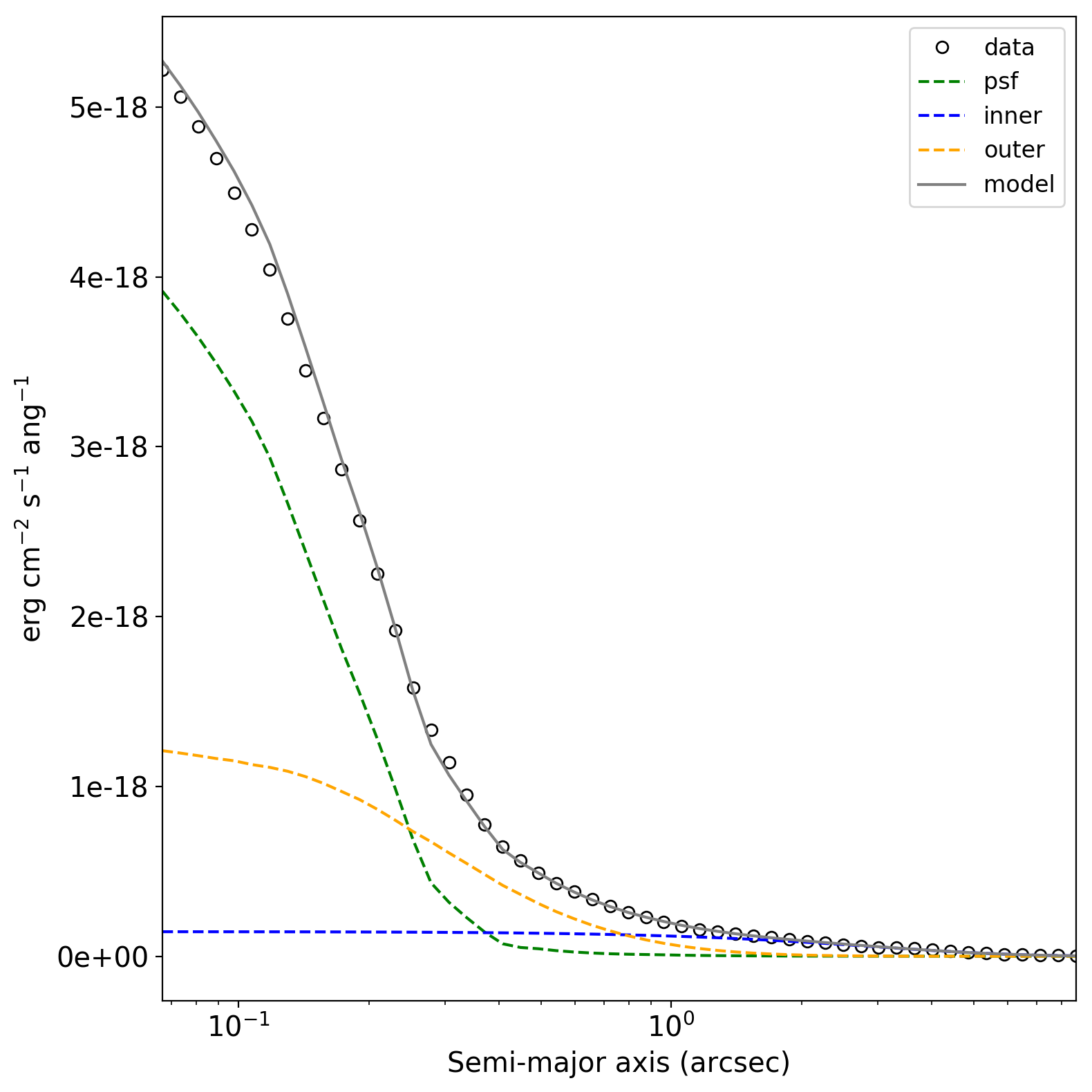}}
\end{minipage}
\caption{Top row: image of RGG 32 in the F110W filter (left); best fit GALFIT model which includes a PSF, inner Sersic component and outer Sersic component (middle); residuals (right). Bottom row: Left panel shows the observed surface brightness profile of RGG 32 with open circles. The best fit model is shown in gray, with the components being shown in green (PSF), blue (inner Sersic) and orange (outer Sersic). The residuals are shown in the lower panel. The right panel shows the average intensity along a given isophote for the data and the intensity as a function of radius. }
\label{fig:RGG32_Model}
\end{figure*}

RGG 32 (Figure \ref{fig:RGG32_Model}) is an Sa galaxy with a dim ring/spiral structure in the disk, most readily seen in the residual panel of Figure \ref{fig:RGG32_Model}. The bulge component has a Sersic index of $n\approx1.6$ and a half-light radius of 0.29 kpc. While the Sersic index of this component is less than 2 it has a round profile and dominates over the disk brightness at inner radii, we therefore classify it as a classical bulge. The disk component has a Sersic index of $n\approx0.75$ with a half-light radius of 2.03 kpc.

\subsection{RGG 48} \label{app:RGG48}

\begin{figure*}
\centering
\subfloat{\includegraphics[width=\textwidth]{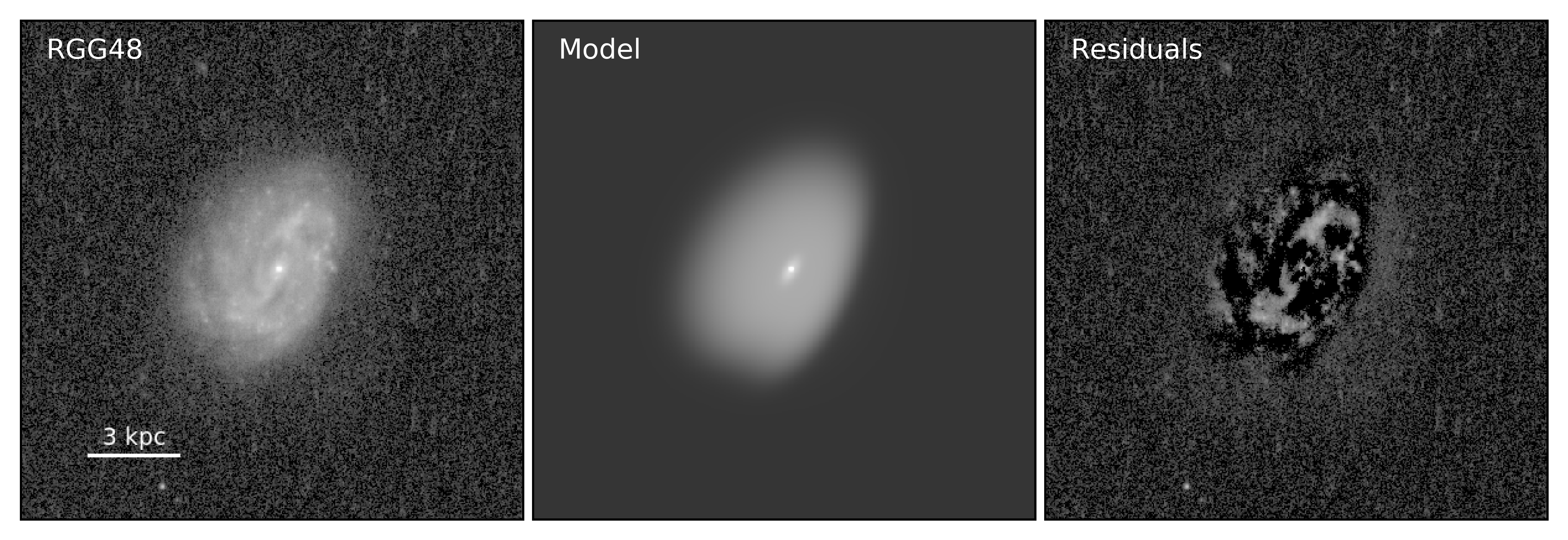}}
\vspace{-0.1cm}
\begin{minipage}{.47\linewidth}
\centering
\subfloat{\includegraphics[scale=0.47]{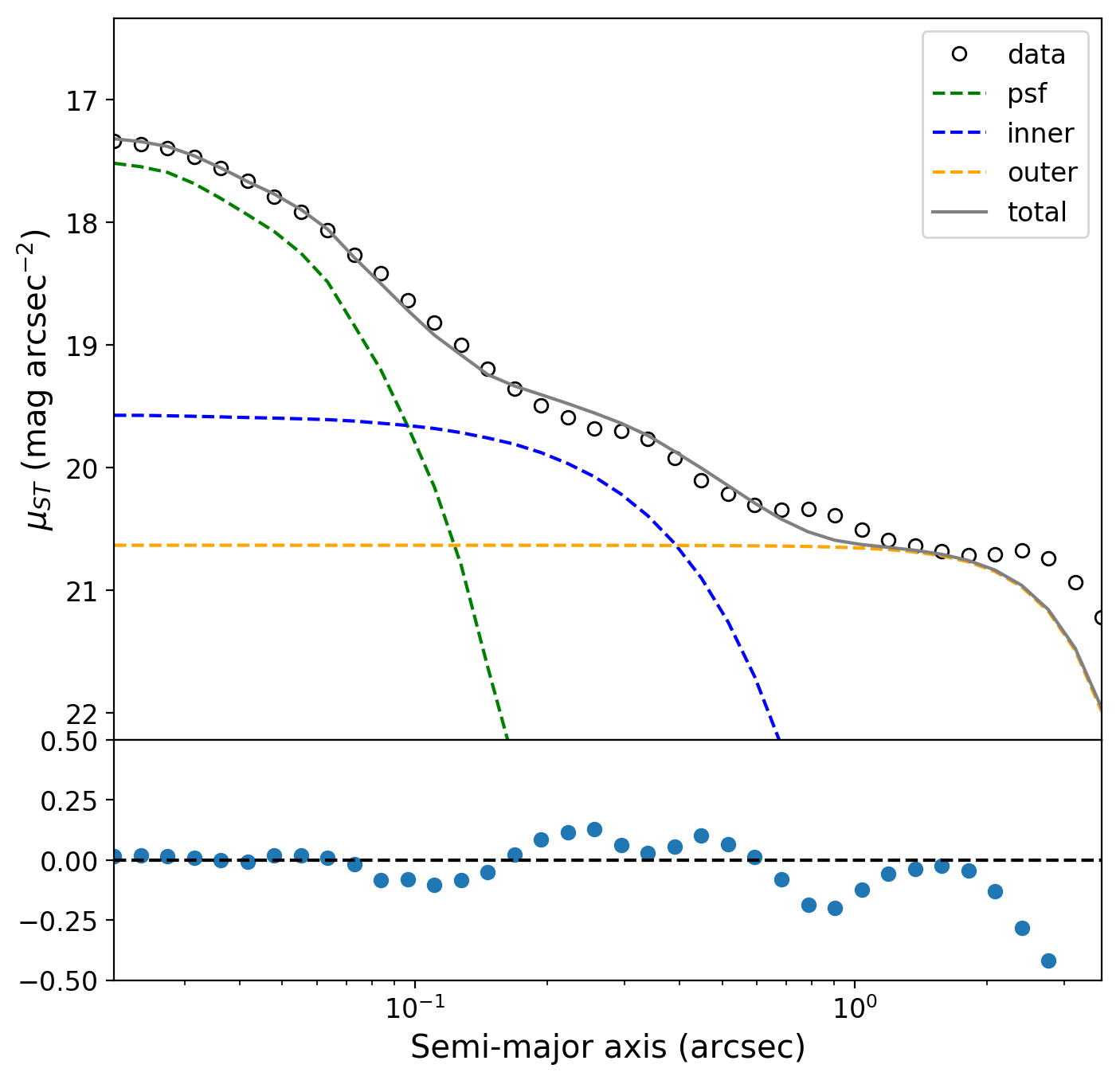}}
\end{minipage}
\hfill
\begin{minipage}{.47\linewidth}
\centering
\subfloat{\includegraphics[scale=0.4]{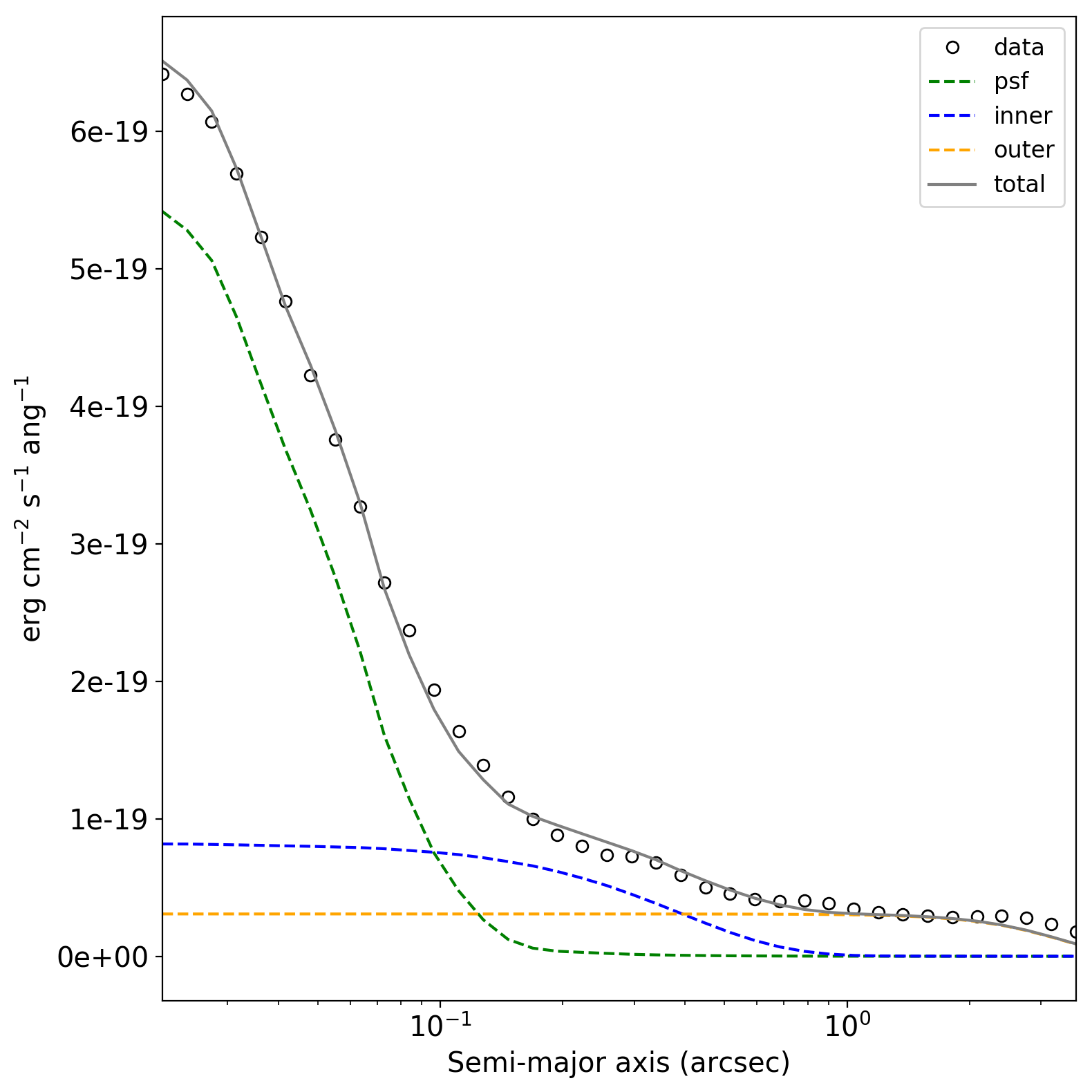}}
\end{minipage}
\caption{Top row: image of RGG 48 in the F606W filter (left); best fit GALFIT model which includes a PSF, inner Sersic component and outer Sersic component (middle); residuals (right). Bottom row: Left panel shows the observed surface brightness profile of RGG 48 with open circles. The best fit model is shown in gray, with the components being shown in green (PSF), blue (inner Sersic) and orange (outer Sersic). The residuals are shown in the lower panel. The right panel shows the average intensity along a given isophote for the data and the intensity as a function of radius. }
\label{fig:RGG48_Model}
\end{figure*}

RGG 48 (Figure \ref{fig:RGG48_Model}) is a disk dominated spiral galaxy with a great deal of structure and star-forming regions. Most clearly seen in the residuals of Figure \ref{fig:RGG48_Model} there is a partially obscured ring around the small central bulge and AGN. This is accompanied by an asymmetric/obscured, barred spiral which is embedded in an asymmetric disk. While these features are readily picked out in the residual image they are actually quite dim and GALFIT has difficulty converging on a model which takes into account more than the inner bulge and the disk. We find the inner component has a Sersic index of $n\approx0.6$ and a half-light radius of 0.3 kpc. The low Sersic index, flat profile and presence of features such a the stellar ring allow this component to be confidently classified as a pseudobulge. The outer component has a Sersic index of $n\approx0.3$ with half-light radius of 2.12 kpc, fitting the classic description of an exponential disk. The disk is interesting as it is quite asymmetric and slightly offset from the pseudobulge component.

\subsection{RGG 119} \label{app:RGG119}

\begin{figure*}
\centering
\subfloat{\includegraphics[width=\textwidth]{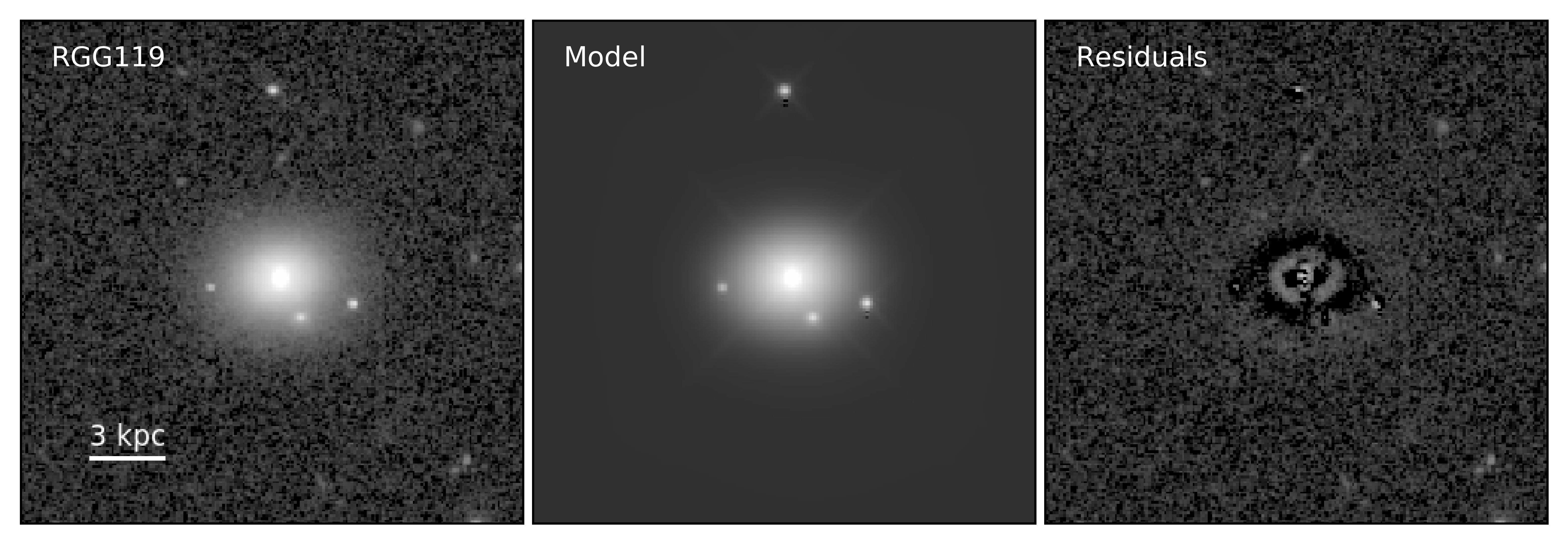}}
\vspace{-0.1cm}
\begin{minipage}{.47\linewidth}
\centering
\subfloat{\includegraphics[scale=0.47]{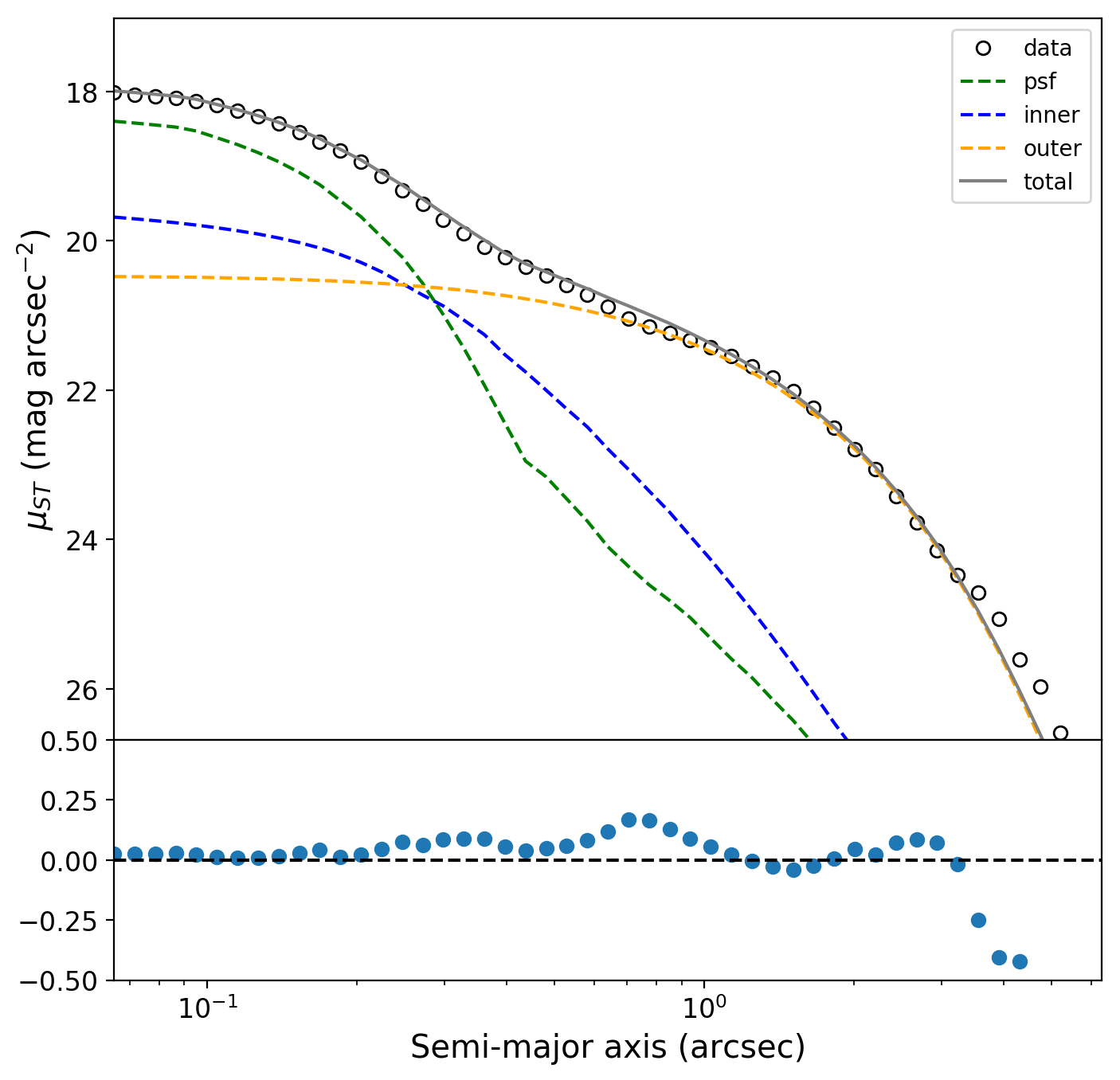}}
\end{minipage}
\hfill
\begin{minipage}{.47\linewidth}
\centering
\subfloat{\includegraphics[scale=0.4]{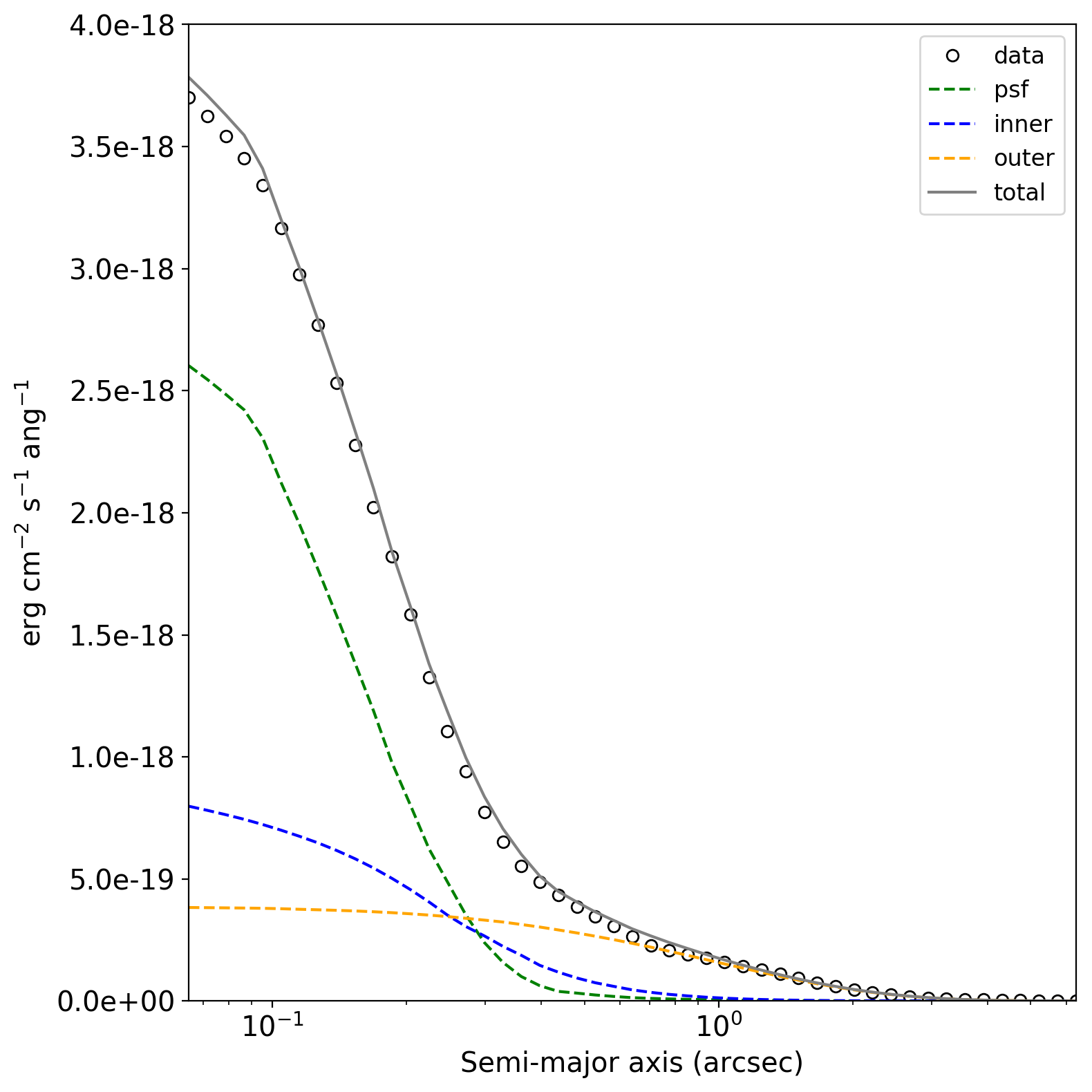}}
\end{minipage}
\caption{Top row: Top row: image of RGG 119 in the F110W filter (left); best fit GALFIT model which includes a PSF, inner Sersic component and outer Sersic component (middle); residuals (right). Bottom row: Left panel shows the observed surface brightness profile of RGG 32 with open circles. The best fit model is shown in gray, with the components being shown in green (PSF), blue (inner Sersic) and orange (outer Sersic). The residuals are shown in the lower panel. The right panel shows the average intensity along a given isophote for the data and the intensity as a function of radius. }
\label{fig:RGG119_Model}
\end{figure*}

RGG 119 (Figure \ref{fig:RGG119_Model}) is an S0 galaxy with evidence of a small stellar ring seen in the residuals. The inner component has a Sersic index of $n\approx 2.5$ with a half-light radius of 0.17 kpc. The relatively high Sersic index, round profile and dominance of the inner component at small radii clearly place this feature in the classical bulge category. The outer component has a Sersic index of $n\approx 0.9$ with half-light radius of 1.02 kpc.

\subsection{RGG 127} \label{app:RGG127}

\begin{figure*}
\centering
\subfloat{\includegraphics[width=\textwidth]{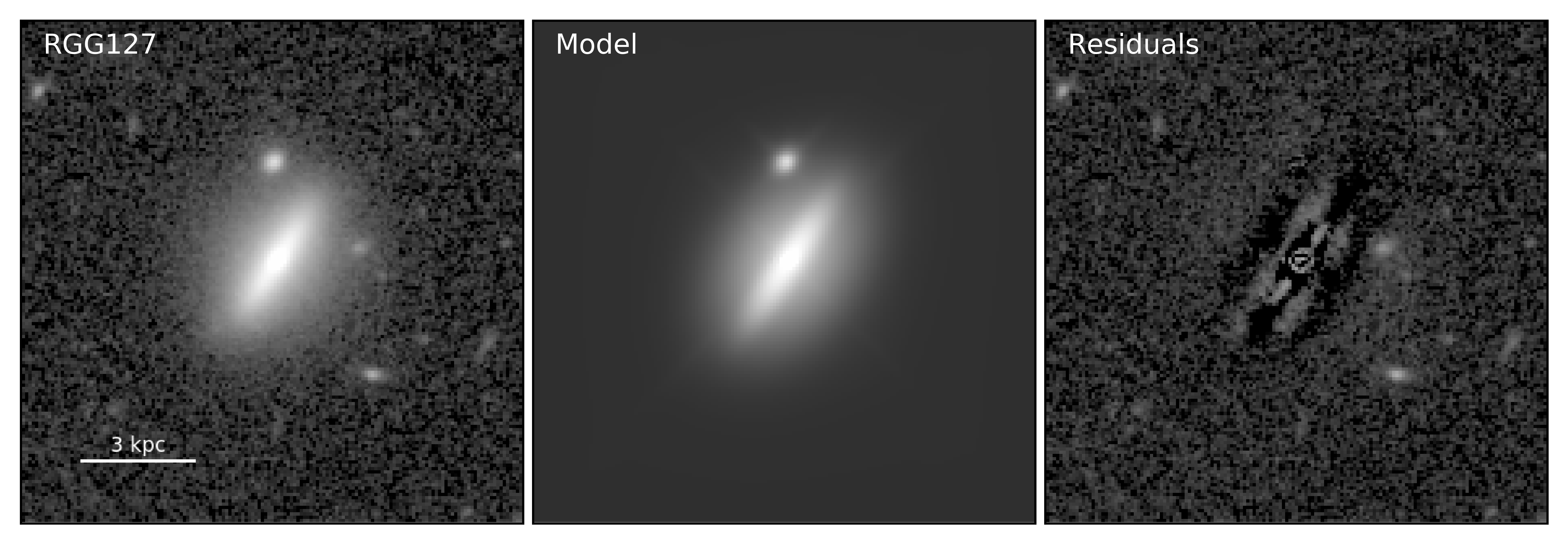}}
\vspace{-0.1cm}
\begin{minipage}{.47\linewidth}
\centering
\subfloat{\includegraphics[scale=0.47]{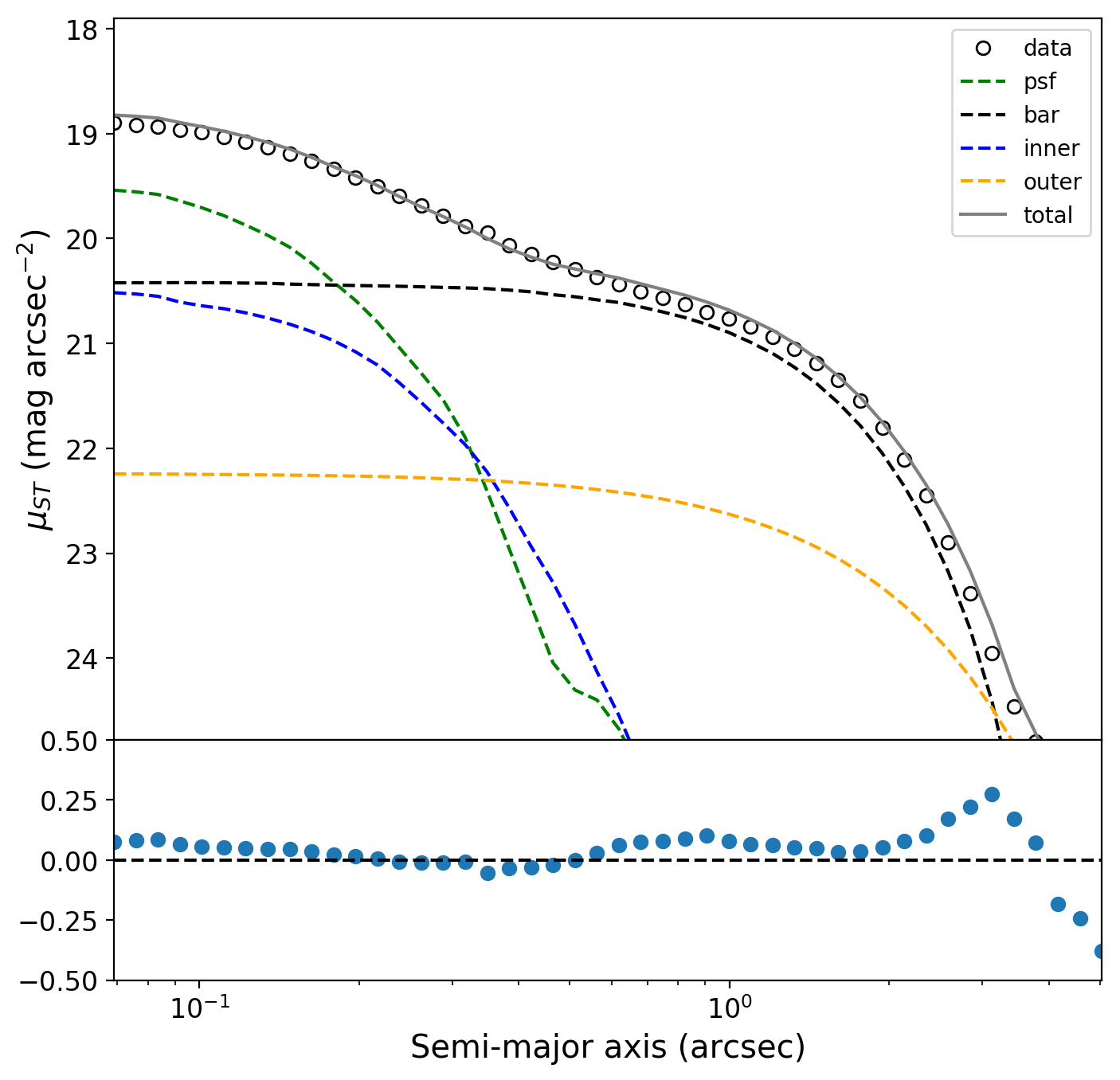}}
\end{minipage}
\hfill
\begin{minipage}{.47\linewidth}
\centering
\subfloat{\includegraphics[scale=0.4]{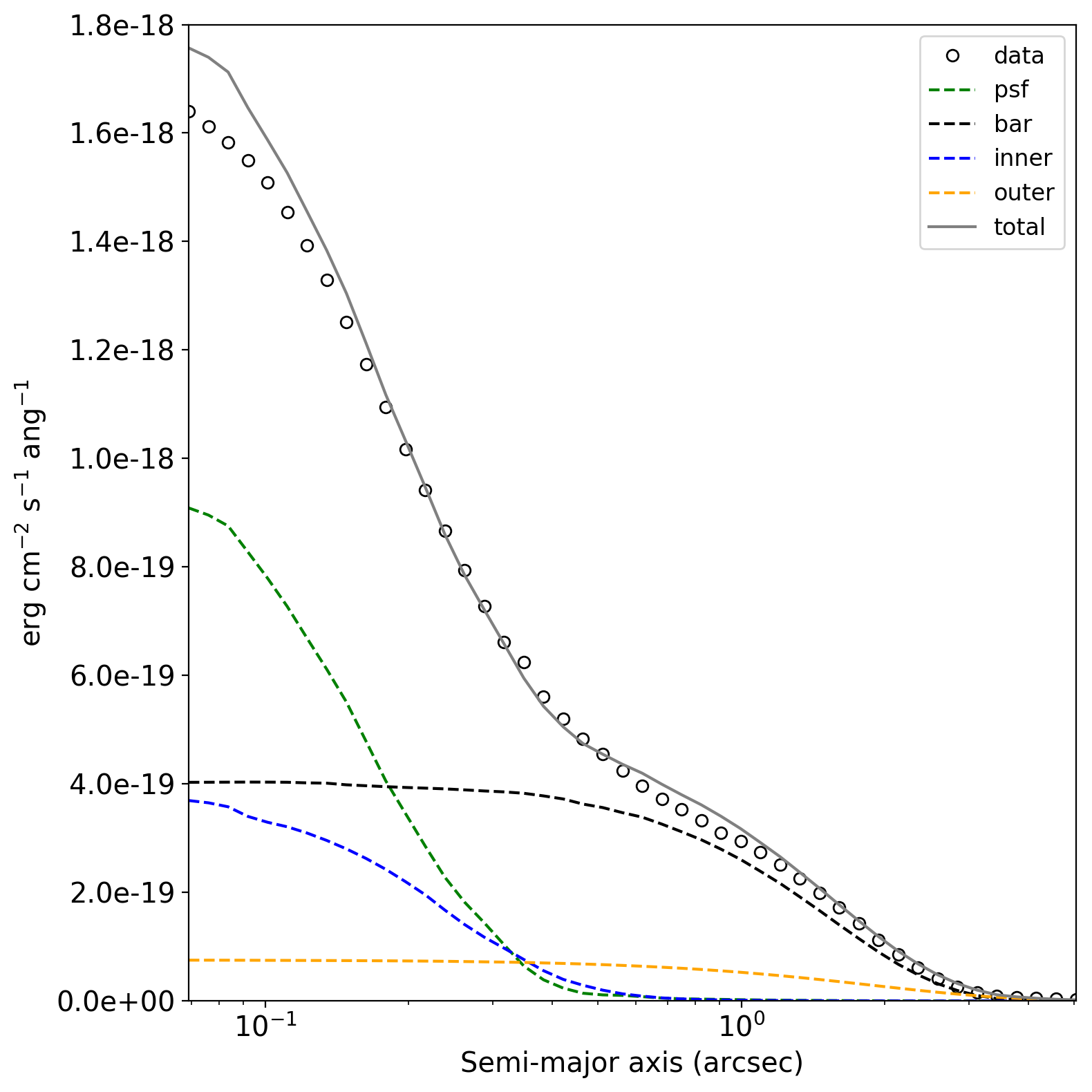}}
\end{minipage}
\caption{Top row: image of RGG 127 in the F110W filter (left); best fit GALFIT model which includes a PSF, inner Sersic component and outer Sersic component (middle); residuals (right). Bottom row: Left panel shows the observed surface brightness profile of RGG 127 with open circles. The best fit model is shown in gray, with the components being shown in green (PSF), blue (inner Sersic), black (bar) and orange (outer Sersic). The residuals are shown in the lower panel. The right panel shows the average intensity along a given isophote for the data and the intensity as a function of radius. }
\label{fig:RGG127_Model}
\end{figure*}

RGG 127 (Figure \ref{fig:RGG127_Model}) is more difficult to classify. Two obvious interpretations of its structure come to mind. First is that we are observing a disk galaxy edge-on, with the bright extended feature being the edge-on disk and the diffuse, rounder feature being an envelope/halo surrounding the disk. The second interpretation would be that the galaxy is being observed close to face-on and the bright, thin feature is a bar and the outer feature is a dimmer disk. In either interpretation the galaxy requires 3 Sersic components to be cleanly fit. The inner most component has a Sersic index of $n\approx1$ with a half light radius of 0.09 kpc. From the low Sersic index of this component, the presence of a possible bar feature and the low central surface brightness indicates this component should be classified as a pseudobulge. The bar structure has a Sersic index of $n\approx0.5$ and a half-light radius of 0.88 kpc. The outer disk/envelope has a Sersic index of 0.7 with a half-light radius of 1.25 kpc, indicating a roughly exponential disk.

\bibliography{ref}

\end{document}